\PassOptionsToPackage{sort&compress}{natbib}

\documentclass[preprint,11pt,]{elsarticle}

\usepackage{graphicx}
\usepackage{amsmath}
\usepackage{booktabs}
\usepackage{subcaption}
\usepackage{amssymb} 
\usepackage{comment}
\usepackage{multirow}
\usepackage{tabularx}
\usepackage{xcolor}
\usepackage{wrapfig}
\usepackage{footmisc}
\usepackage{sidecap}

\usepackage[letterpaper,margin=1in]{geometry}
\usepackage[numbers,sort&compress]{natbib}

\usepackage{hyperref}
\hypersetup{
    colorlinks=true,    
    linkcolor=blue,     
    citecolor=blue,     
    urlcolor=blue       
}

\begin{document}

\begin{frontmatter}

\title{Precision Tokenization for Multi-Modal Data in High Energy Physics}

\renewcommand*{\thefootnote}{\fnsymbol{footnote}}

\author[slac]{Jeffrey Krupa\footnote{Equal Contributions, \\ 
\phantom{xxx} Corresponding Authors: 
\href{mailto:jkrupa@slac.stanford.edu}{jkrupa@slac.stanford.edu} , 
\href{mailto:samklein@slac.stanford.edu}{samklein@slac.stanford.edu},
\href{mailto:nilotpal.kakati@weizmann.ac.il\\ }{nilotpal.kakati@weizmann.ac.il}}}
\author[slac]{Samuel Klein\footnotemark[1]}
\author[weiz]{Nilotpal Kakati\footnotemark[1]}
\author[tum]{Daniel Murnane}
\author[mpm]{Baran Hashemi}
\author[weiz]{\\ Eilam Gross}
\author[tum]{Lukas Heinrich}
\author[slac]{Michael Kagan}

\affiliation[slac]{organization={SLAC National Accelerator Laboratory}, city={Menlo Park}, state={CA}, country={USA}}
\affiliation[weiz]{organization={Weizmann Institute of Science}, city={Rehovot}, country={Israel}}
\affiliation[tum]{organization={Technical University of Munich}, city={Munich}, country={Germany}}
\affiliation[mpm]{organization={Max Planck Institute for Mathematics in the Sciences}, city={Leipzig}, country={Germany}}

\begin{abstract}
Foundation models have recently gained significant attention in fields such as high energy physics (HEP).
The tokenizers that convert detector data into discrete inputs for such models, however, are comparatively unexplored.
This paper addresses this gap by learning discrete and compressed tokenized representations of HEP datasets.
We demonstrate the approach on four modalities covering two different simulated detectors that vary in complexity. 
Treating tokenization as a rate--distortion problem, a systematic hyperparameter scan yields four design rules for tokenizing HEP data that hold across all set-based modalities studied.
The cost of tokenization is evaluated on tasks including jet $p_T$ regression, particle flow reconstruction, and flavor tagging.
We find that tokenization preserves much of the information needed for these tasks at compression factors of approximately 4--11$\times$ while keeping the jet $p_T$ resolution below 1\%.
We also present \texttt{heptokens}, an open-source library for tokenizing HEP data.
\end{abstract}

\begin{keyword}
vector quantization  \sep foundation models \sep tokenization \sep jet tagging  \sep high energy physics \sep multi-modal data
\end{keyword}

    \end{frontmatter}
\renewcommand*{\thefootnote}{\arabic{footnote}}
\setcounter{footnote}{0}

\clearpage
\section{Introduction}
\label{sec:intro}
\begin{wrapfigure}{r}{0.45\textwidth}
    \raggedleft
    \includegraphics[width=0.45\textwidth]{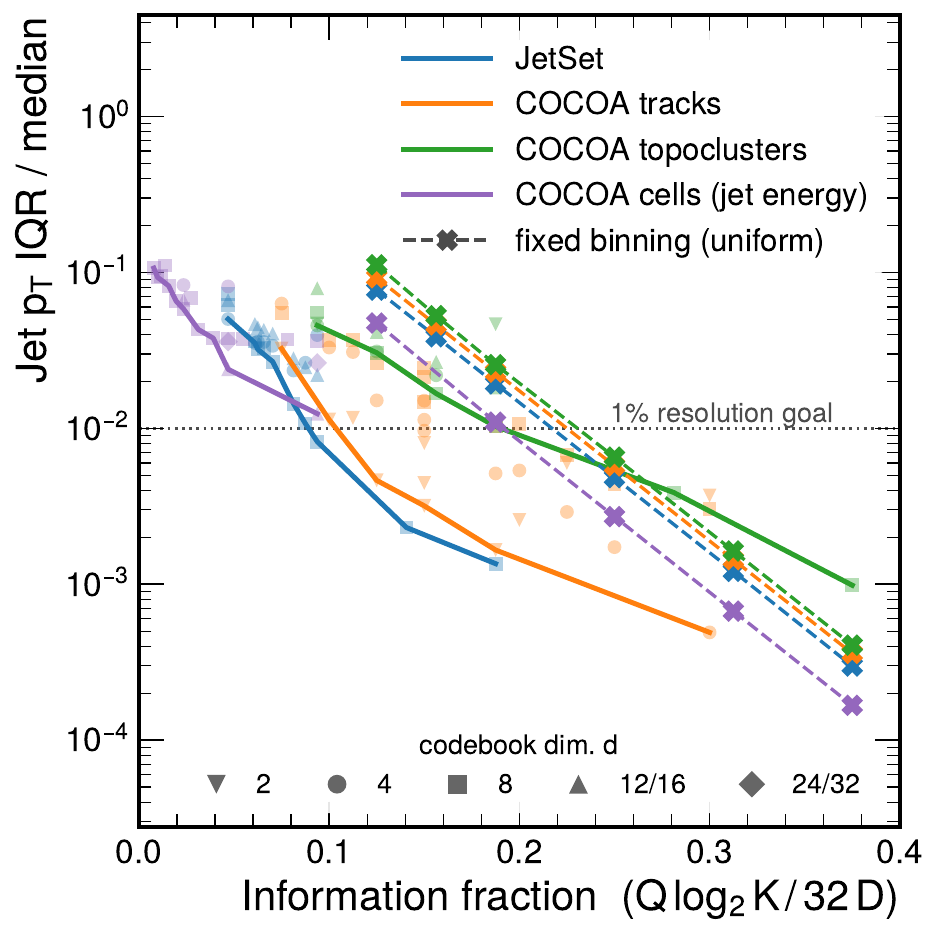}
    \caption{Pareto frontier of jet $p_T$ resolution versus information fraction $Q\log_2 K/(32D)$ for the JetSet and COCOA tokenizers.
    Marker shape denotes the codebook dimension $d$ ranging from 2--32.
    Faded points are individual $(K,d,Q)$ configurations, and solid lines trace the frontier for each tokenizer.
    Dashed lines show a baseline that bins each feature independently into uniform bins, without learning.
    JetSet, COCOA track and COCOA topocluster tokenizers reach the 1\% resolution goal at compression factors of approximately 4--11$\times$, while COCOA calorimeter cell tokenizers remain slightly above the target, reaching 1.24\% resolution at a compression factor of 11$\times$.
    All set-based tokenizers quantize the detector coordinates together with the other features (positions-in,~\autoref{sec:design_rules}). 
    Cell tokenization is described in~\autoref{sec:cell_loss}.
    }
    \label{fig:rate-distortion}
\end{wrapfigure}

Progress in natural language processing and computer vision has been accelerated by the development of \emph{foundation models}, which we define as models that are pre-trained on large corpora of data and then adapted to multiple downstream tasks.
A common ingredient in these models is the representation of the inputs as sequences of discrete \emph{tokens} drawn from a finite vocabulary.
This paradigm is native to text, and has seen successes in images~\cite{vqvae,dalle,llamagen}, audio~\cite{soundstream,encodec}, and video~\cite{videopoet}.
For continuous domains, the appeal of tokenization is that it maps sparse local structure to compact indices, reduces modeling to classification, compresses the data volume by orders of magnitude, and enables diverse modalities to share a single architecture and training recipe whose performance scales predictably with data and compute~\cite{scalinglaws,chinchilla}.

The state of the art in high energy physics (HEP) operates directly on continuous inputs~\cite{particlenet,particletransformer,GN2,GN3}.
The community has begun exploring tokenization~\cite{Golling:2024abg,OmniJet,Visive:2025flm,Visive:2026rzu} and scaling~\cite{ATL-SOFT-PUB-2026-002,vigl2026neural,amram2026neuralscalinglawsjet,uslu2026,bahl2026scalinglawsamplitudesurrogates}, but whether tokenization is necessary, especially in multi-modal settings, remains an open question.
Continuous generative decoders have been shown to outperform tokenized objectives at current scale~\cite{Leigh_2025}, although the comparison has not yet been made in the large-data regime and in multi-modal settings where the scaling and uniform representation advantages of discrete sequences are expected to emerge.

In this paper we perform a comprehensive study of how HEP data can be tokenized and quantify what it costs.
We show how to explore the Pareto-frontier of performance versus tokenization-driven compression, as displayed in~\autoref{fig:rate-distortion} and discussed in Section~\ref{sec:frontier}.
Learned tokenizers reach a 1\% jet $p_T$ resolution with up to $2.4\times$ fewer bits than a non-learned per-feature binning.
We distill the result into four design rules for tokenizing HEP data (Section \ref{sec:design_rules}).
In brief: quantize in several successive stages to refine residual differences, use a large vocabulary of codes, match the dimensionality of the codes to the modality and to the bit budget, and quantize only what is modality-specific, encoding global quantities such as detector coordinates in a fixed binning instead.
These rules hold across all set-based modalities studied.
We measure what tokenization costs downstream by evaluating tokenized inputs on tasks including jet $p_T$ regression, particle flow reconstruction~\cite{hep4m}, and flavor tagging~\cite{particlenet,GN2}, measuring performance as a function of compression.
The tokenizers, scans, and evaluations are provided as \textit{heptokens}\footnote{\label{note1}Available at \url{https://github.com/treasure-AmSC/heptokens}.}, an open-source library for building tokenized representations of particle physics data.

The inputs to HEP models vary from sensor-level sparse calorimeter deposits to variable-length sets of reconstructed particles.
We use residual vector-quantized variational autoencoders (RVQ-VAEs, referred to as RVQ in the following)~\cite{vqvae,soundstream}, which replace each element by $Q$ integer codes from which the element is reconstructed by the decoder.
Only the input features and encoder--decoder differ between modalities.
We train tokenizers on four detector modalities of differing complexity: charged-particle tracks from the ATLAS flavor-tagging open dataset \emph{JetSet}~\cite{jetset}, and tracks, topoclusters, and calorimeter cells from the \emph{COCOA} dataset~\cite{COCOA}.

\section{Related work}
\label{sec:related}
Tokenization underpins the first attempts at HEP foundation models. 
Masked particle modeling~\cite{Golling:2024abg} pre-trains by masking constituents in a set and predicting the missing elements, using the discrete codes of a pre-trained VQ-VAE as targets.
That work includes a limited study of the VQ-VAE itself, and reported difficulty preventing codebook collapse, in which a handful of code vectors absorb all assignments.
Our work extends the VQ-VAE approach with residual layers, which is significantly more robust to codebook collapse.
A subsequent study~\cite{Leigh_2025} found that continuous reconstruction targets can outperform tokenized objectives for training foundation models, which leaves open how much physics a discrete representation retains, and at what compression.

Tokenization also enables multi-modal foundation models for particle physics.
Kakati \textit{et al.}~\cite{hep4m} represent every modality in a jet as sets of discrete tokens.
A single model is trained on these tokens to map any subset of these modalities onto any other.
Particle flow, detector simulation, and charged-energy subtraction then become applications of the same model in different directions.

OmniJet-$\alpha$~\cite{OmniJet} tokenizes jet constituents with a VQ-VAE, building on the tokenization from masked particle modeling, and demonstrates transfer between an unsupervised generation task and a supervised tagging task.
Subsequent work~\cite{OmniJet2} refines the pre-training strategy, combining masked and next-token objectives.
In these models the tokenizer is an ingredient adopted to enable a backbone, rather than the object of study.

While the majority of efforts in this direction have focused on jets, tokenization efforts have been extended to other modalities, like calorimeter showers~\cite{OmniJet-c} and detector hits for tracking~\cite{Huang:2024voo}.
Most recently, SPADE~\cite{SPADE} tokenizes high-granularity calorimeter data by embedding each feature of a hit independently, avoiding the information loss of a single token.
In each case the tokenizer serves a single downstream generative or reconstruction model.
In contrast, this work is the first in HEP to treat tokenization as a rate--distortion problem, formalizing the trade-off between compression and reconstruction fidelity. 
We survey it systematically across tracks, topoclusters, and calorimeter cells in two datasets, quantifying the fidelity of physical observables as a function of compression.
\section{Methods}
\label{sec:method}

\label{sec:overview}
All detector modalities are represented as variable-length sets of elements, but they differ substantially in cardinality and spatial structure. 
Sparse modalities such as tracks and topoclusters comprise at most a few tens of objects per event, and each reconstructed object is treated directly as one element $x_i\in\mathbb{R}^D$. 
The calorimeter can contain orders of magnitude more deposits and can be represented as an image~\cite{Cogan_2015}, such that the cell can be partitioned into fixed-size image patches~\cite{esser2020taming}.
An encoder $f_\theta$ maps the sample to per-element latents in $\mathbb{R}^d$.
Two types of tokenization are considered, differing in the choice of encoder--decoder model.
In \emph{no-context tokenization}, each $x_i$ is encoded independently by an MLP model. 
In \emph{in-context tokenization}, a Transformer~\cite{vaswani2023attentionneed} encodes each $x_i$ with attention over all other elements in the sample.
Since in-context tokenization outperforms no-context tokenization at every tested configuration, Transformer encoder--decoder models are used throughout this paper. 
For example, the largest Transformer and MLP models achieve jet $p_T$ resolutions of 0.82\% and 1.27\%, respectively (\ref{app:model_dataset_size}).

\begin{figure}
    \centering
    \includegraphics[width=0.95\linewidth]{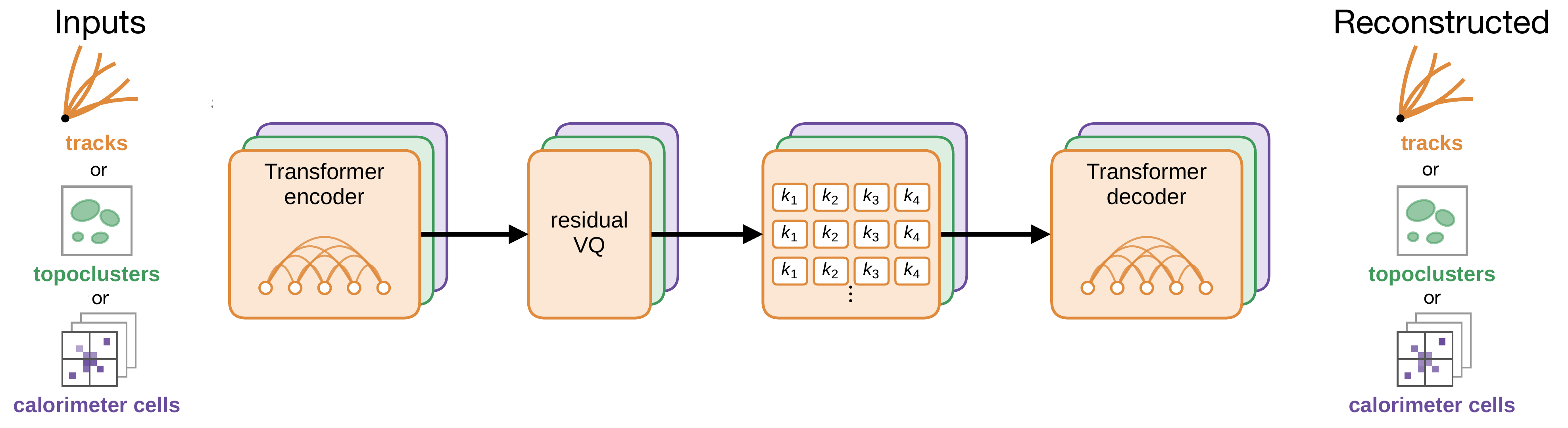}
    \caption{A schematic of the tokenizer architecture.
    Each modality (tracks, topoclusters, or calorimeter cells) has its own Transformer encoder, residual VQ, codebooks, and Transformer decoder.
    Tokenizers for each modality are trained separately, shown in stacked boxes.
    The encoders and decoders use self-attention across all elements of a given modality in an event.
    The RVQ maps each element to $Q$ discrete tokens $k_1,\ldots,k_Q$.}
    \label{fig:modalities}
\end{figure}
The encoder, quantizer, and decoder are trained jointly by minimizing a reconstruction loss against $\{x_i\}$, together with a commitment loss.
After training, a sample of $N$ valid elements is represented by its $N\times Q$ integer tokens for codebooks with $K$ entries.
A separate tokenizer is trained for each modality, as shown in~\autoref{fig:modalities} for tracks, topoclusters, and calorimeter cells illustrated for $Q=4$.
This compact discrete representation can be decoded back to the input or used directly for downstream tasks such as reconstruction, regression, and classification. 
This recipe is modality-agnostic and only the input features and the encoder--decoder model differ between detector objects. 
Detector coordinates $(\eta,\phi)$ are either included as input features and encoded by the RVQ alongside the other quantities (``positions-in'' configuration), or discretized into $K_{\textrm{pos}}$  bins per coordinate and appended as separate token indices after the object tokens (``positions-out'' configuration)~\cite{hep4m}.

\subsection{Residual vector quantization}
\label{sec:rvq}

Each element latent $z_e\in\mathbb{R}^{d}$ is discretized by a residual vector quantizer with $Q$ stages. 
Stage $q$ has a codebook $\{e_{q,k}\}_{k=1}^{K}$ of $K$ vectors in $\mathbb{R}^{d}$ and quantizes the running residual $r_q$ to its nearest code,
\begin{equation}
    k_q = \arg\min_{k}\, \lVert r_q - e_{q,k}\rVert^2,
    \qquad r_1 = z_e,\qquad r_{q+1} = r_q - e_{q,k_q},
\end{equation}
so that the quantized latent is the sum of the selected codes,
\begin{equation}
    z_q = \sum_{q=1}^{Q} e_{q,k_q}.
    \label{eq:z_q}
\end{equation}
Each element is thus encoded by the $Q$ integer indices $(k_1,\dots,k_Q)$ and successive stages refine the residual, so that $Q$ stages of size $K$ span up to $K^{Q}$ code combinations while storing only $Q\times K$ vectors. 
During training, gradients reach the encoder through the straight-through estimator~\cite{ste}, and the codebooks are updated by an exponential moving average of the latents assigned to each code rather than by a gradient term. 
The tokenizer is trained end-to-end to minimize the total objective $\mathcal{L}$:
\begin{equation}
\mathcal{L} = \mathcal{L}_\mathrm{rec} + \beta\,\mathcal{L}_\mathrm{commit}, \qquad
    \mathcal{L}_\mathrm{commit} = \frac{1}{Q}\sum_{q=1}^{Q}
      \big\lVert r_q - \mathrm{sg}[e_{q,k_q}]\big\rVert^2 ,
\end{equation}
with $\mathrm{sg}[\cdot]$ the stop-gradient operator and the loss is averaged over the $Q$ stages and over valid elements. The commitment loss $\mathcal{L}_\mathrm{commit}$ keeps the encoder outputs close to their selected codes.
The reconstruction loss $\mathcal{L}_\mathrm{rec}$ is an $L_1$ error on the input features for the set-based modalities, and the mixed energy/occupancy objective of Section~\ref{sec:cell_loss} for calorimeter cells. 
The weight $\beta$ controls the strength of the commitment term. 
The codebook size $K$, code dimension $d$, and number of stages $Q$ are the quantizer's three hyperparameters.
How to choose $(K,d,Q)$ for a given HEP modality, and the associated cost in fidelity, is addressed in Section~\ref{sec:results}.

\subsection{Cell reconstruction objective}
\label{sec:cell_loss}

\begin{figure}
    \centering
    \includegraphics[width=\linewidth]{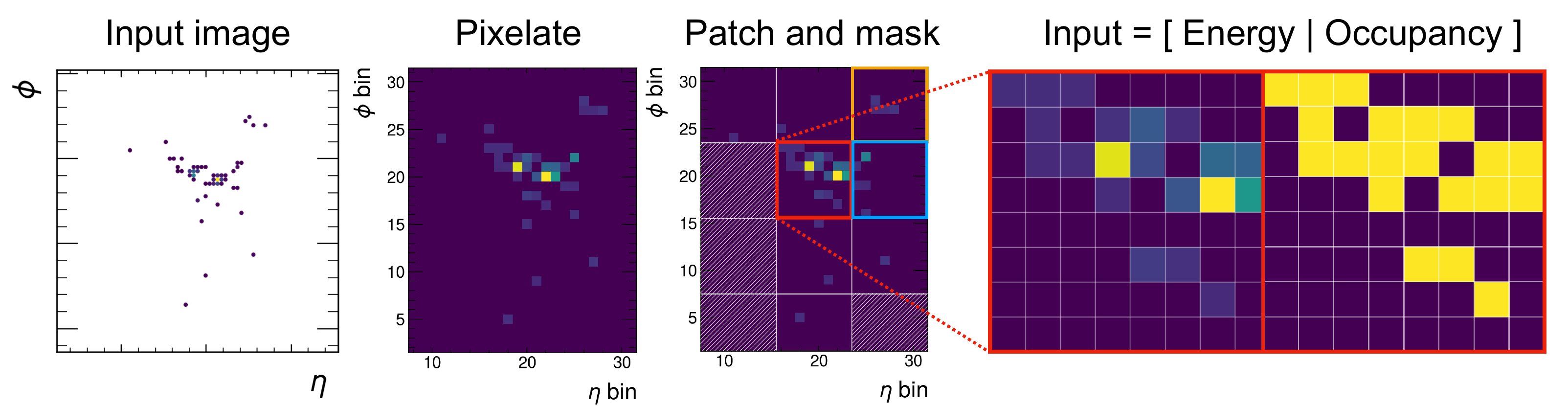}
    \caption{A schematic displaying how calorimeter cells are tokenized.
    Cells in each layer are pixelated into an $(\eta,\phi)$ image and tiled into $8\times8$ patches.
    Patches with no energy are masked out.
    For each patch that remains, the pixel energies and a binary occupancy map, $o=[E>0]$, are concatenated to form the input.
    The patches of an event are encoded jointly by a Transformer, quantized into $Q$ indices per patch by the RVQ, and decoded back to energy and occupancy. 
    }    

    \label{fig:pixelize_and_patchify}
\end{figure}

Unlike the track- and topocluster-based tokenizers, which operate on unordered sets, the cell tokenizer operates on a fixed-geometry image.
Cells could also be tokenized as a point cloud~\cite{OmniJet-c}, which avoids a voxel grid, at the cost of encoding coordinates that carry no information beyond which cell fired.
An image encodes them in the pixel index instead.
A schematic of this is shown in~\autoref{fig:pixelize_and_patchify}.
Cells are binned into per-layer $(\eta,\phi)$ images and tiled into $8\times 8$-pixel patches.
Every patch carries the compressed pixel energies and a binary occupancy indicator, $o=[E>0]$, marking pixels with non-zero deposited energy.
To handle the sparsity of the image, empty patches are flagged by a patch-level mask.
Each patch is treated as one element and encoded as $Q$ indices.
The encoder and decoder are Transformers whose self-attention layers let all $n_\mathrm{patch}$ patches of an event exchange information, with empty patches masked out and each patch's $(\eta,\cos\phi,\sin\phi)$ position added as a positional encoding.
In this scheme, each patch's reconstruction depends on the whole event, not just the pixels belonging to the patch.

The training loss combines the cell reconstruction terms with the RVQ commitment loss,
\begin{equation}
\mathcal{L} = \underbrace{\sum_{i:\,o_i=1} w_i\,\bigl(\hat{y}_i - y_i\bigr)^2
 + \mathrm{BCE}\!\left(\hat{o},o\right)}_{\text{reconstruction}}
 + \beta\,\mathcal{L}_\mathrm{commit},
 \qquad
 w_i = \frac{E_i^{\alpha}}{\sum_{j:\,o_j=1} E_j^{\alpha}},
\end{equation}
where $\hat{y}$ is the decoder’s prediction of the compressed pixel energy $y = E^{0.3}$, and $E$ is the linear cell energy.
The energy term is evaluated only on truth-occupied pixels ($o=1$) and weighted with $\alpha=0.4$ to emphasize the high-energy cells that dominate the jet energy.
The occupancy binary cross-entropy (BCE) is averaged over all pixels of non-empty patches.
Because the energy and occupancy are predicted jointly but the energy loss ignores truth-empty pixels, the network is free to predict arbitrary energy values where $o=0$.
At inference the sparse reconstructed cells are obtained by applying a sigmoid to the occupancy logits and retain only pixels with predicted probability of being non-empty greater than 50\%.
The energy predicted elsewhere is zeroed. 
The tokenizer must therefore learn both \emph{in which} cells energy is deposited, and \emph{how much} energy is deposited. 
The procedure is illustrated step by step in~\ref{app:calo}.

\section{The \texttt{heptokens} framework}
\label{sec:software}

The methods described here are implemented in \texttt{heptokens}\footref{note1}, an open-source package for learning discrete tokenizations of heterogeneous detector data. 
This library implements the RVQ of Section~\ref{sec:method}.
Encoders, decoders, and data modules share common abstract interfaces, so a new class of detector object is supported by adding a data module and, where the reconstruction target requires it, a decoder head, rather than a new training pipeline.

The framework is built on PyTorch Lightning and configured through Hydra config groups for model, data, and callbacks, so architectures and hyperparameters are declarative and can be overridden from the command line. 
The RVQ interface exposes the codebook size, code dimension, and number of residual stages as first-class hyperparameters.
Studies that proceed in multiple steps, such as preprocessing, training, tokenizing, evaluation, downstream tasks, are orchestrated as reproducible Snakemake workflows~\cite{snakemake} in Pixi environments~\cite{pixi}.
Trained models export their token indices and codebooks, yielding a compact discrete representation of an event that can be reused for downstream tasks such as classification, generative modeling, or foundation-model pre-training. 

\section{Datasets}
\label{sec:datasets}

We study four detector modalities drawn from two datasets of differing realism.
The COCOA simulation~\cite{COCOA} provides tracks, topoclusters, and calorimeter cells from the same single-jet events, while the ATLAS open dataset JetSet~\cite{jetset} provides charged-particle tracks from $t\bar{t}$ jets with flavor labels.
The datasets, along with data preprocessing strategies, are summarized in~\ref{app:datasets}.

The three set-based modalities (COCOA tracks, COCOA topoclusters, and JetSet tracks) share the same preprocessing.
Transverse momenta (and, for topoclusters, the energy, cluster width $\rho$, and ECAL/HCAL energies) are transformed as $x \to \log(1+x)$, and every feature is then standardized to zero mean and unit variance.
The standardization is fit separately for each modality on a training subsample ($10^{6}$ jets for JetSet, $10^{5}$ events for COCOA) and applied only to valid elements, leaving padded slots untouched.
Calorimeter cells are handled differently, as described in Section~\ref{sec:cocoa}.

\subsection{COCOA data}
\label{sec:cocoa}
 
A separate tokenizer is trained on each of the three COCOA modalities.
\emph{Tracks} are charged particles propagated through the $3.8$~T magnetic field
and smeared to detector-level resolution rather than reconstructed from hits.
Each track is described by $p_T$, $\eta$, $\phi$, and the transverse and longitudinal impact parameters $d_0$ and $z_0$. 
There are up to $15$ tracks per event. 
\emph{Topoclusters} are groups of calorimeter cells formed by a topological
clustering algorithm~\cite{ATLAStopo}, with the signal-to-noise thresholds
configured as in~\cite{COCOA}. 
Each is described by the position $\eta,\phi$, the total energy $E$ and its electromagnetic and hadronic components $E_\mathrm{ECAL},E_\mathrm{HCAL}$ and the cluster width $\rho$ and shape $\sigma_\eta,\sigma_\phi$, with up to $50$ clusters per event. 
Both modalities are variable-length sets, padded to a fixed length with a validity mask, and the tokenizer is trained purely to reconstruct the inputs. 
\emph{Calorimeter cells} are populated by the energy deposited by jets across the six layers of a sampling calorimeter.
Each cell provides a position $(\eta,\phi)$, a layer index, and a deposited energy.
Each pixel of the per-layer images (Section \ref{sec:cell_loss}) holds $\sum_{\mathrm{cell}} E_\mathrm{cell}^{\,p}$ over the cells in that bin, with a power-law rescaling $p=0.3$ used to regularize the cell-energy spectrum, which spans many orders of magnitude.
The coarsest layer uses a smaller $4\times 4$ patch, which is further zero-padded to $8\times 8$ so that patches from all layers have a common length of $64$.
This yields a total of $156$ patches per event across the six layers.

\subsection{ATLAS flavor-tagging open data}
\label{sec:atlas}

Jets from the open ATLAS dataset JetSet~\cite{jetset} are categorized into four classes by hadron-based truth flavor: light, charm ($c$), bottom ($b$), and tau ($\tau$). 
Each jet is an unordered set of up to $40$ charged-particle \emph{tracks}, with $20$ features per track: the transverse momentum $p_T$ and its fraction of the jet $p_T$; the angular offsets from the jet axis $\Delta\eta,\Delta\phi$; the transverse and longitudinal impact parameters ($d_0$, $z_0\sin\theta$, and
their beamspot-relative variants) together with their uncertainties and lifetime-signed significances;
and the track angle $\theta$ and curvature $q/p$. 
Four jet-level kinematic features ($p_T$, mass,
$\eta$, $\phi$) are retained for reconstruction diagnostics. 
A small fraction of jets (0.015\%) are excluded because they contain at least one misreconstructed constituent with $p_T>1~\mathrm{TeV}$.

\section{Tokenization results}
\label{sec:results}

Each tokenizer is scanned over codebook size $K$, code dimension $d$, and number of residual quantizers $Q$.
The training-set sizes differ per modality.
COCOA track tokenizers are trained on 20~M events, COCOA topocluster tokenizers are trained on 5~M events, COCOA calorimeter cell tokenizers are trained on 1.6~M events, and JetSet tokenizers are trained on 7~M events.
Dataset-size scans at representative configurations show that the track, topocluster, and JetSet tokenizers do not improve beyond these sizes (\ref{app:model_dataset_size}).
Calorimeter cell events are much larger, so the cell scan uses 1.6~M events for computational reasons.
All metrics are evaluated on held-out test samples.

Fidelity is judged at the constituent level by the mean absolute error (MAE) of each feature, averaged over features, and at the jet level by the resolution of the jet $p_T$ rebuilt from the reconstructed constituents.
This resolution is the interquartile range (IQR), the difference between the 75th and 25th percentiles of the $p_T^\mathrm{reco}/p_T^\mathrm{truth}$ distribution, divided by its median.
Here, $p_T^\mathrm{reco}$ and $p_T^\mathrm{truth}$ are the $p_T$ of the jet rebuilt from the same constituents after and before tokenization, respectively.
The IQR therefore measures the error introduced by the tokenizer alone.
We target a resolution below 1\%, making the tokenization subdominant to jet energy resolution of the ATLAS and CMS detectors, which range from 5--25\%~\cite{atlas_jer,cms_jer}.
For calorimeter cells, the analogous quantity is the IQR/median of $E^\mathrm{reco}/E^\mathrm{truth}$, where the jet energy is the sum of the cell energies before and after tokenization.

\subsection{Rate--distortion frontier}
\label{sec:frontier}

We measure reconstruction fidelity versus tokenization-driven compression, tracing the Pareto frontier for each modality.
An element (a track, a topocluster, or an $8\times8$ cell patch) with $D$ continuous 32-bit features carries $32D$ bits.
After tokenization it is encoded as $Q$ indices into codebooks of size $K$, carrying $Q\log_2K$ bits.
The information fraction preserved by tokenization is therefore $I = Q\log_2K/(32D)$.

The scan of tokenizer models for all modalities is summarized in~\autoref{fig:rate-distortion}.
Only positions-in configurations are shown. 
Positions-out configurations are compared in~\autoref{sec:design_rules}.
Each faded point is one tokenizer trained at a single $(K,d,Q)$ configuration, placed at its information fraction $I$ on the horizontal axis.
The reconstructed jet $p_T$ resolution is shown on the vertical axis.
Color denotes the separate tokenizers for each modality: JetSet (blue), COCOA tracks (orange), COCOA topoclusters (green), and COCOA calorimeter cells (purple).
Marker shapes denote the code dimension $d$.
For each modality, the solid line traces the Pareto frontier, defined as the best resolution reached at a given $I$.
Points above the frontier spend as many bits but reach a worse resolution.
The same scan evaluated with the normalized MAE metric is shown in~\ref{app:model_performance_studies}.
Repeated trainings with different seeds give a spread of about 0.02\% in jet $p_T$ resolution, well below the variation across the $(K,d,Q)$ scan.

The learned tokenizers are compared to a baseline that bins each feature independently into uniform bins, shown as dashed lines.
Where both are available, the learned tokenizers reach a better MAE at every $I$, and a better jet $p_T$ resolution at every $I$ except for topoclusters above $I \approx 0.25$.

Three of the four tokenizers reach the 1\% jet $p_T$ resolution target, at different values of $I$.
JetSet tokenizers reach the resolution goal at $I=0.09$, while COCOA tokenizers reach the resolution goal at $I=0.125$ and $I=0.28$ for tracks and topoclusters, respectively.
COCOA calorimeter cell tokenizers remain slightly above the target, reaching a jet energy IQR/median of 1.24\% at $I=0.09$, relative to dense 32-bit pixels.
These correspond to compression factors ranging from approximately 4--11$\times$, depending on modality.
For each modality, we select the point with the lowest $I$ on the Pareto frontier that meets the 1\% target, with one exception for tracks discussed below, and the best point for calorimeter cells.
The selected configurations are listed in~\autoref{tab:hyperparams}.

Jet $p_T$ resolution is sensitive only to the kinematic features.
For COCOA tracks, the most compressed configurations on the frontier reconstruct $p_T$, $\eta$ and $\phi$ well, but lose the impact parameters $d_0$ and $z_0$, which do not enter the jet four-vector.
The normalized MAE, which weights all features equally, shows this directly: no $d=2$ configuration appears on its frontier, which moves to $d=8$ at higher bit rates, the smallest scanned code dimension at least as large as the five track features.
We therefore select $(4096,8,3)$ for COCOA tracks, at $I=0.23$ (4.4$\times$ compression) and 0.7\% jet $p_T$ resolution.

\begin{table}[th]
\centering
    \caption{The selected RVQ tokenizer configurations per modality.
    This is the point with the lowest $I$ on each Pareto frontier that meets the 1\% jet $p_T$ resolution target (except COCOA tracks, selected on MAE) together with the resulting compression.
    Backbones are Transformer encoder--decoders of two sizes, large ($d_\mathrm{model}{=}256$, 6 layers, 8 heads) and medium ($d_\mathrm{model}{=}128$, 4 layers, 8 heads), with the number of trainable parameters in parentheses.
    $K$, $d$, and $Q$ are the codebook size, code dimension, and number of residual stages.
    Bits are the number of bits per element, $Q\log_2 K$, and compression is quoted relative to the $32D$ bits of the $D$ continuous input features.
    Resolution is the IQR/median of $p_T^\mathrm{reco}/p_T^\mathrm{truth}$ for the jet rebuilt from tokenized constituents. 
    Calorimeter cells show the best jet energy resolution $E^\mathrm{reco}/E^\mathrm{truth}$ of the scan$^{\dagger}$.
    Utilization is the fraction of codebook entries in use, averaged over the $Q$ stages.
    }
    
\label{tab:hyperparams}
\small
\setlength{\tabcolsep}{5pt}
\renewcommand{\arraystretch}{1.2}
\begin{tabularx}{\textwidth}{@{} >{\raggedright\arraybackslash}X l c c c c c c c @{}}
\toprule
Modality & Backbone & $K$ & $d$ & $Q$ & Bits & Compr. & Res. & Util. \\
\midrule
Jet constituents (JetSet)  & Transformer L (9.6\,M)  & $32768$ & $8$ & $4$ & 60 & $10.7\times$ & $0.82\%$  & $77\%$ \\
Tracks (COCOA)             & Transformer M (1.6\,M)  & $4096$  & $8$ & $3$ & 36 & $4.4\times$  & $0.68\%$ & $71\%$ \\
Topoclusters (COCOA)       & Transformer M (1.6\,M)  & $4096$  & $8$ & $6$ & 72 & $3.6\times$  & $0.39\%$ & $99\%$ \\
Calorimeter cells (COCOA)  & Transformer L (9.6\,M)  & $4096$  & $16$ & $16$ & $192$ & $10.7\times$ & $1.24\%^{\dagger}$   & 88\% \\
\bottomrule
\end{tabularx}
\end{table}

At the selected configurations, each input feature is reconstructed without visible bias, and the cell tokenizer reproduces the active-cell multiplicity within 0.01\% (shown for COCOA in~\ref{app:model_performance_studies}).
Supporting codebook utilization plots are collected in \ref{app:codebook_utilization} .

\subsection{Recommendations for tokenizing HEP data}
\label{sec:design_rules}

\begin{figure}
    \centering
    \includegraphics[width=0.9\linewidth]{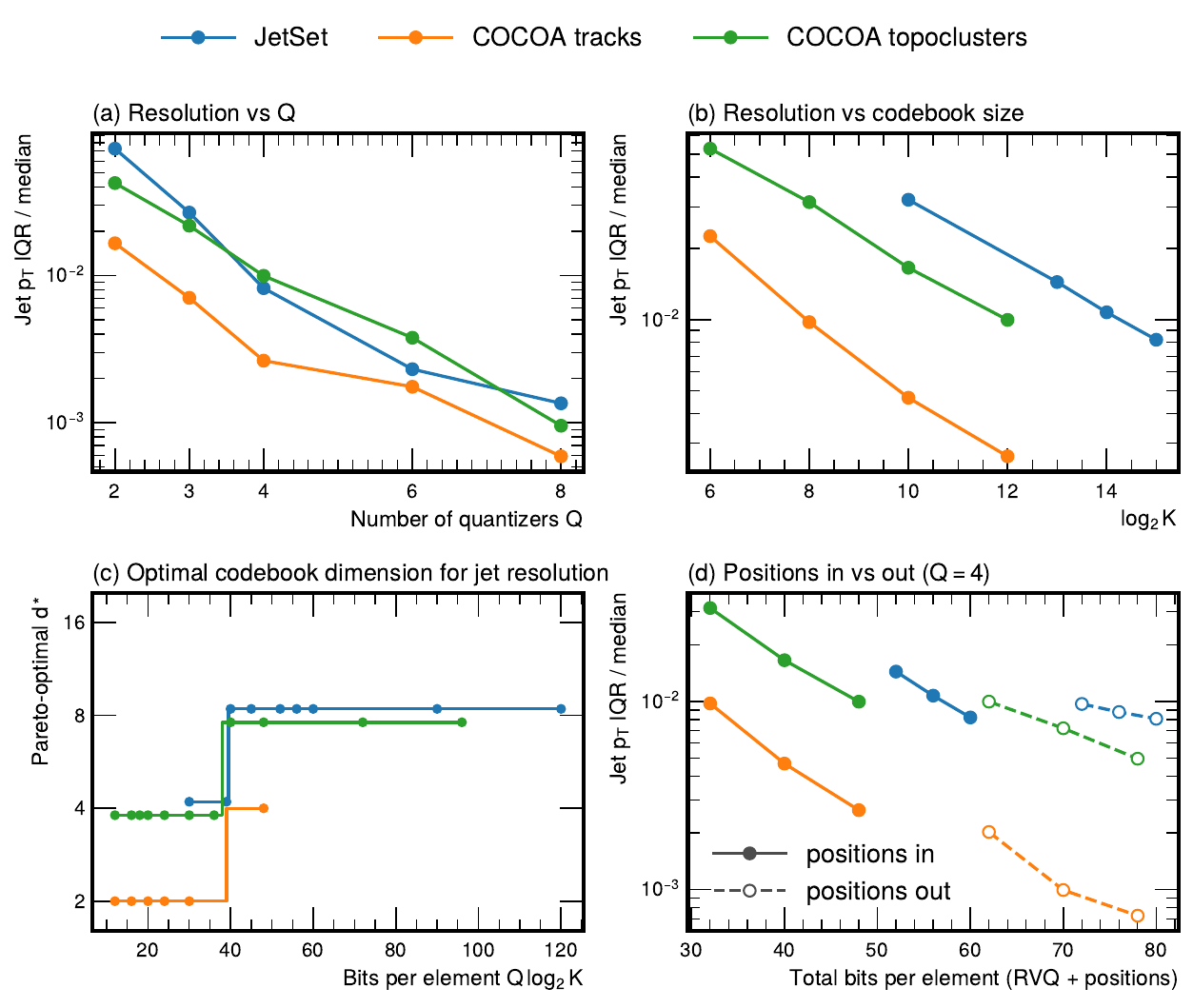}
    \caption{Summary of the four design rules.
    Jet $p_T$ resolution for JetSet (blue), COCOA tracks (orange), and COCOA topoclusters (green).
    \textit{(a)} Resolution versus number of quantizers $Q$ at fixed $(K,d=8)$ for each modality.
    \textit{(b)} Resolution versus codebook size $K$ at fixed $(d=8,Q=4)$.
    \textit{(c)} Codebook dimension that is Pareto-optimal for jet resolution $d^*$ versus bit budget $Q\log_2 K$.
    \textit{(d)} Resolution versus total bits per element with positions in the RVQ (filled circles) or positions encoded in a fixed binning with $K_{\mathrm{pos}}=1024$ outside of the RVQ (open circles), adding 20 (30) position bits for JetSet (COCOA).
    }
    \label{fig:four_rules}
    \end{figure}

The hyperparameter scans support four rules that hold across the set-based modalities and datasets investigated in this paper, each corresponding to a panel in~\autoref{fig:four_rules}.

\begin{enumerate}[(\alph*)]
    \item Reconstruction improves with the number of quantizers $Q$, since each stage refines the residual of the previous one.
    A single quantizer stage does not converge in this setting and has a tendency to collapse to a small subset of available codes.
    \item Reconstruction improves with codebook size $K$, since more codes provide finer binning of the latent space.
    \item The code dimension $d$ should scale with both the bit budget and the intrinsic dimensionality of the target observable.
    At low bit budgets the Pareto-optimal $d$ for jet resolution ($d^*$) is small, and it grows as the budget increases as shown in \autoref{fig:rate-distortion}.
    When including other parameters, such as impact parameters, the codebook dimension must get larger. 
    In practice, tracks and topoclusters are best served by $d=8$ and calorimeter cells by $d=8$--$16$.
    \item Encoding $(\eta,\phi)$ in a fixed global binning outside the RVQ (positions-out) improves resolution at fixed RVQ bit-rate, at the cost of $\log_2 K_{\mathrm{pos}}$ bits per coordinate.
    Counting these bits, positions-out gives a net gain when coordinates make up a large fraction of the element’s features, as for COCOA tracks, while for topoclusters and JetSet tracks positions-in reaches the same resolution with fewer total bits.  
    Independent of rate, a fixed binning encodes positions at a known resolution and provides a position vocabulary shared across modalities.
\end{enumerate}
    
\section{Evaluation tasks}
\label{sec:cocoa_downstream}

We explore how well tokenized representations preserve physics level knowledge, testing $p_T$ regression and particle flow reconstruction in COCOA, and jet-flavor classification in JetSet.
The particle flow task additionally probes whether independently tokenized tracks and topoclusters can be composed jointly, without any joint tokenizer training.

\begin{wrapfigure}[31]{r}{0.45\textwidth}
    \centering
    \includegraphics[width=\linewidth]{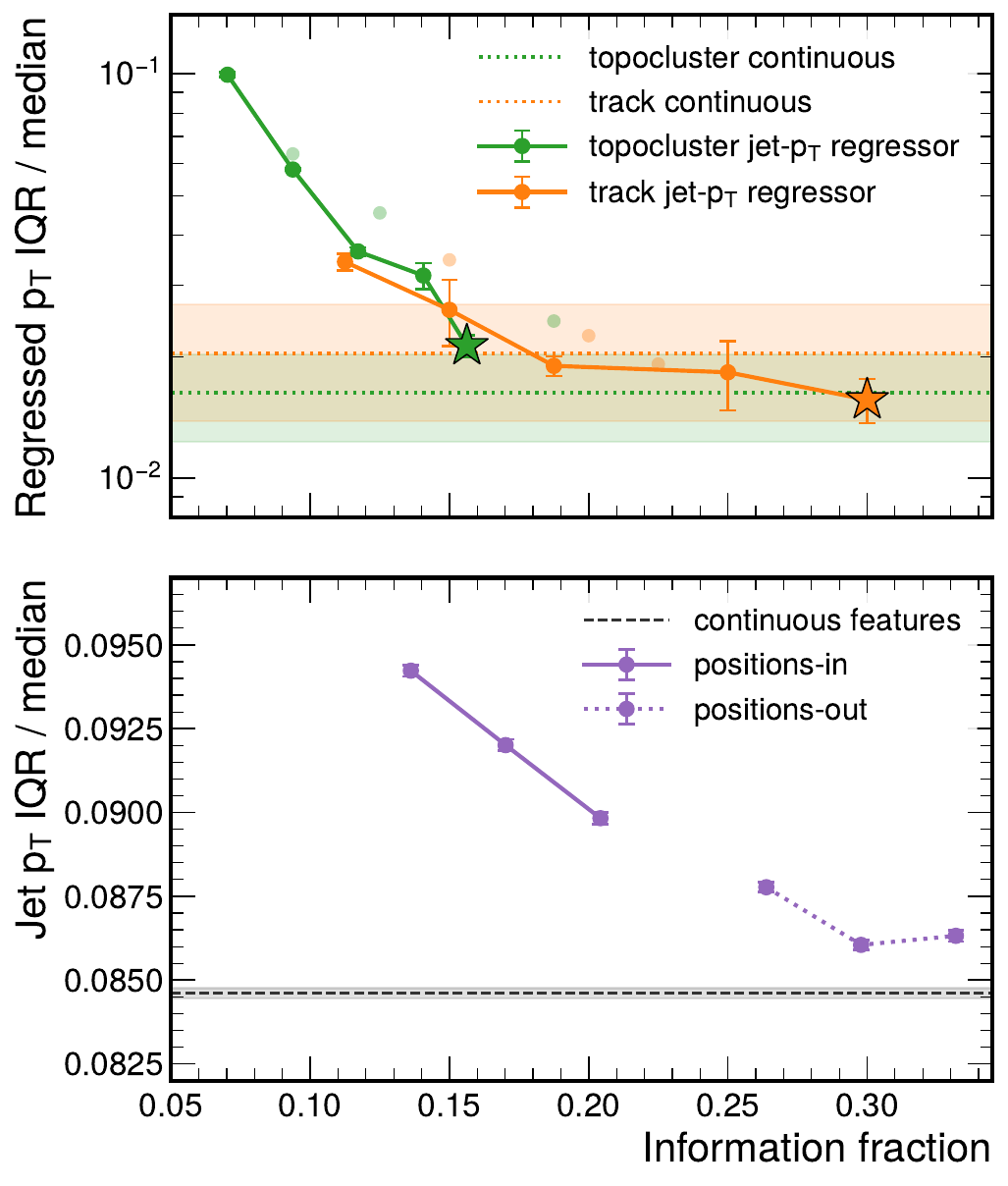}
    \caption{Downstream jet $p_T$ resolution versus total information fraction, including RVQ and position bins.
    \textit{Top}: IQR/median of the regressed-to-true jet-$p_{T}$ ratio for regressors trained on tokenized COCOA tracks (orange) and topoclusters (green), with their continuous-feature baselines (dotted).
    Solid lines trace the Pareto frontier.
    Stars mark the best regression result for each modality.
    Error bars (shaded bands for the baselines) show the standard error of the mean over three independently trained regressors.
    \textit{Bottom}: IQR/median of the particle flow reconstructed-to-true jet-$p_T$ ratio for positions-in (solid) and positions-out (dotted) tokenizers, and the parameter-matched continuous baseline (dashed).
    Error bars are computed by bootstrapping.}
    \label{fig:cocoa_pf_regression}
\end{wrapfigure}

\vspace{0.3cm}
\noindent\emph{Regression in COCOA:} We train jet $p_{T}$ regressors on the tokenized COCOA track and topocluster constituents to predict the jet $p_T$ before tokenization, and quantify performance by the IQR/median of the regressed-to-true jet $p_{T}$ ratio.
The regressor is a Transformer ($d_\mathrm{model}=128$, 4 layers, 8 heads, sum pooling) that consumes the quantized latents $z_q$ without decoding them, trained on 1~M events.
Tokenizers are scanned over $K\in\{64,256,1024,4096\}$ and $Q\in\{3,4\}$ at fixed $d=4$ for tracks and $d=8$ for topoclusters, with three regressor seeds per configuration.
A continuous baseline is trained on the original features with the same regressor.
As shown in~\autoref{fig:cocoa_pf_regression} (top), the resolution improves with $I$.
Stars mark the best regression result for each modality.    
At these points, the track regressor matches its continuous-feature baseline within uncertainties, and the topocluster regressor lies just above its baseline.

\vspace{0.3cm}
\noindent\emph{Particle flow in COCOA:} We further evaluate the tokenizers on a particle flow reconstruction task~\cite{hep4m}.
A Transformer (15~M parameters) receives the track and topocluster tokens and predicts the tokens of the truth particles.
The truth particle tokens encode $p_{T}$, $\eta$, $\phi$, $E$, and a class (charged particle, neutral hadron, or photon).
Each input object is represented by $Q$ consecutive tokens, one per residual-quantization level. 
In the positions-out variant, $(\eta,\phi)$ are removed from the learned RVQ tokenizer, and the RVQ tokens of each object are followed by three position tokens ($\eta$, $\cos\phi$, $\sin\phi$) from a fixed uniform binning with $K_{\textrm{pos}}=1024$.
Each input token is mapped to the model space by an 8-dimensional per-token vector followed by a shared linear projection.

The predicted tokens are decoded by a fixed truth-particle tokenizer ($K=128$, $Q=3$ with jet $p_T$ resolution 0.8\%), and jets are formed by summing the decoded particle four-vectors.
We scan $I$ by varying $K$ at $Q=4$, training each model on $88$~M events.
To compute $I$ for the combined input, we use the mean number of features per input element, weighting the track (5) and topocluster (8) feature counts by their mean multiplicities per event (3.62 and 12.95), which gives $D = 7.35$.
For positions-out, the $3 \times 10$ bits of the position tokens are added to the numerator, $I = (Q \log_2 K + 30)/(32D)$, so positions-out uses 30 more bits per element than positions-in at the same $(Q,K)$.
As shown in~\autoref{fig:cocoa_pf_regression} (bottom), with positions-in the jet-$p_{T}$ resolution improves with $I$, from $\mathrm{IQR}/\mathrm{median}=0.094$ to $0.090$.
With positions out of the RVQ and encoded in a fixed binning instead, the resolution reaches $0.086$ at $I=0.30$, including the position bits.
The figure also shows a baseline, labeled ``continuous features'', in which the same Transformer model receives the continuous track and topocluster features instead of tokens, mapped to its 384-dimensional latent space by a two-layer MLP, and predicts the same truth-particle tokens.
This baseline reaches $\mathrm{IQR}/\mathrm{median}=0.0845$.
With positions out, tokenizing the inputs therefore costs as little as $\approx2\%$ (relative) in resolution, at $I=0.30$.
Unlike the regression target, the particle flow truth is the generator-level particle set, so the absolute resolution in the lower panel is not comparable to the panel above.

\vspace{0.3cm}
\noindent\emph{Classification in JetSet:}
\label{sec:classification}
We test the information retained by the JetSet tokenizers with a jet-flavor classification task.
Unlike the particle flow model, which embeds token indices, the classifier receives the summed codebook vector $z_q$ for each track, probing the bottleneck of compressing a track into an 8-dimensional vector.
The classifier (11~M trainable parameters) embeds its inputs into a 512-dimensional space, followed by self-attention layers and a flavor-classification head.
We train it on the tokens of four positions-in tokenizers along the Pareto frontier, $(K,Q)\in\{(2^{10},3),(2^{10},4),(2^{13},4),(2^{15},4)\}$, and compare to an identical model given the 20 raw track features, preprocessed as in~\autoref{sec:datasets}.
All classifiers are trained on 140~M jets and evaluated on a held-out set of 30~M jets.

\begin{figure}[htbp]
    \centering
    
    \begin{subfigure}[b]{0.31\textwidth}
\includegraphics[width=\textwidth]{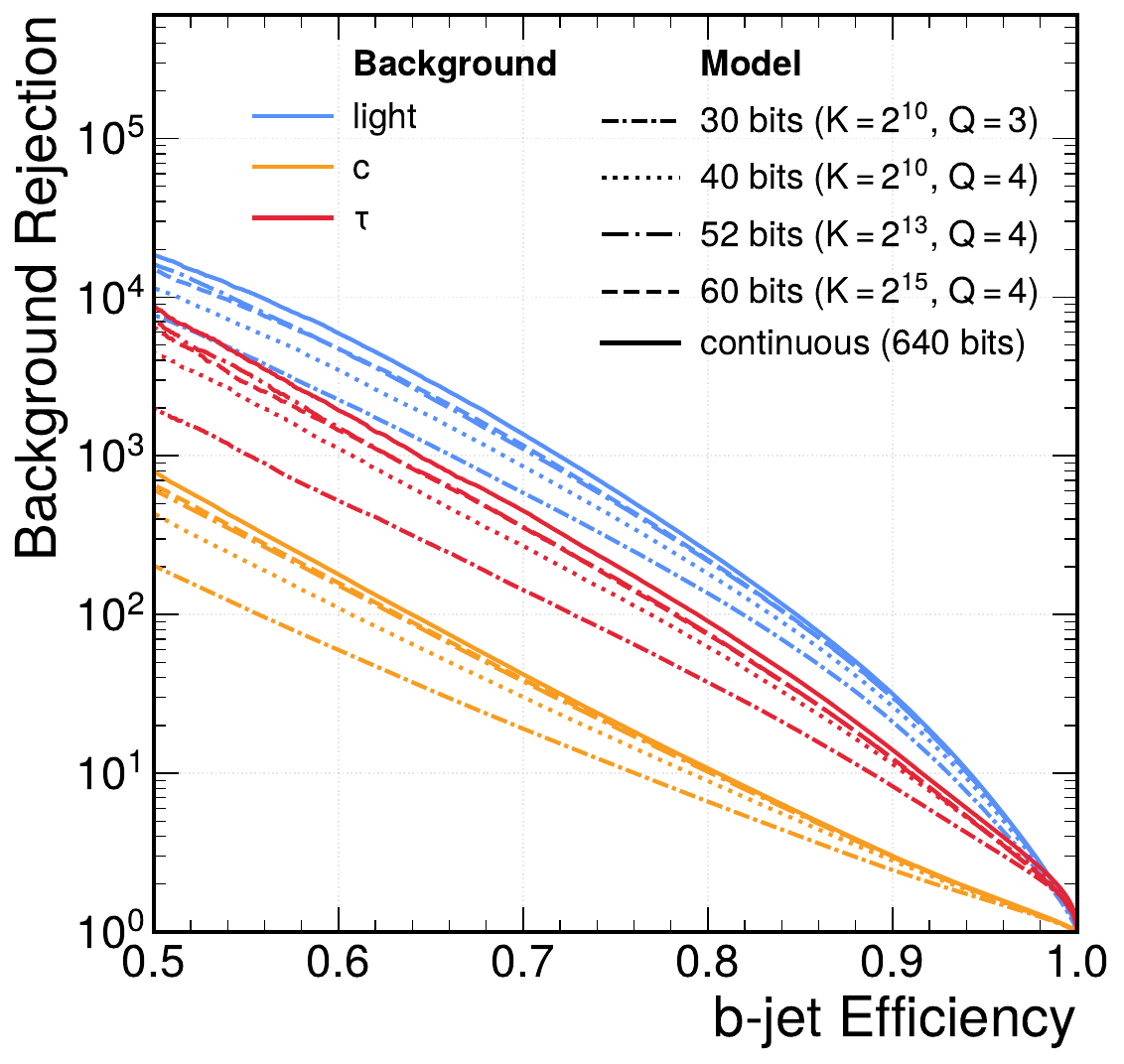}
    \caption{}
    \label{fig:roc_btag_a}
    \end{subfigure}
    \hfill
    \begin{subfigure}[b]{0.35\textwidth}
    \includegraphics[width=\textwidth]{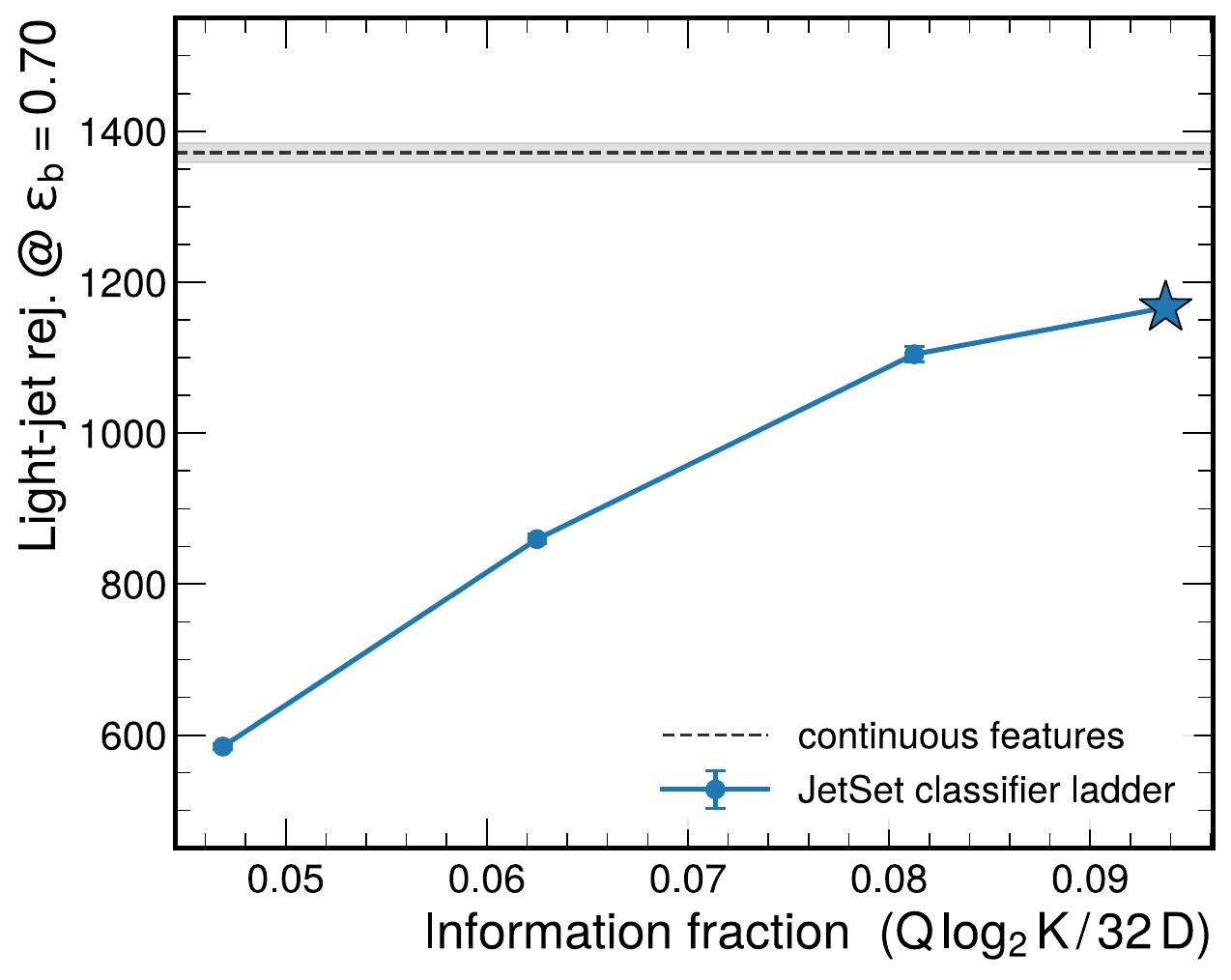}
        \caption{}
        \label{fig:roc_btag_b}
    \end{subfigure}
    \hfill
    \begin{subfigure}[b]{0.21\textwidth}
    \includegraphics[width=\textwidth]{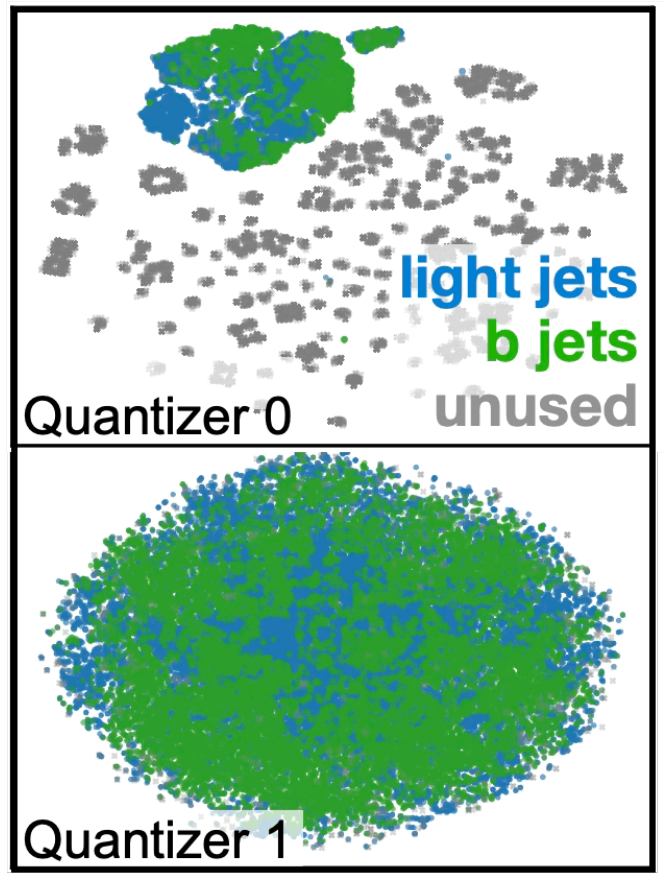}
    \caption{}
    \label{fig:roc_btag_c}
    \end{subfigure}
    
    \caption{
    Performance of jet tagging on JetSet.
    \textit{a)}: $b$-tagging ROC curve showing light/c/$\tau$-jet rejections versus $b$-jet efficiency.
    Rejection of each flavor is shown as a different color.
    The solid curve shows a training on continuous features, while the dashed curves show various RVQ settings corresponding to different $(K,Q)$ combinations with positions-in, as shown in the legend.
    \textit{b)}: light-jet rejection at $\varepsilon_b = 0.70$ as a function of the information fraction.
    \textit{c)}: two-dimensional t-SNE~\cite{tsne} embedding of the codebook vectors of the selected JetSet tokenizer. 
    Each code is a point, colored by the majority flavor of the jets whose constituents are assigned to it.
    Codes with a $c$ or $\tau$ majority are too few to be visible. 
    } 
    \label{fig:roc_btag}
\end{figure}

The results are summarized in~\autoref{fig:roc_btag_a}.
Tokenization preserves most, but not all, of the discriminating information, with measurable losses in the high-rejection tails. At 60 bits per element, the tokenized classifier retains 73--93\% of the continuous rejection, and the loss grows as bits are removed (\autoref{fig:roc_btag_b}, \autoref{tab:classification}).
A t-SNE embedding of the JetSet codebooks in~\autoref{fig:roc_btag_c} shows what the stages learn: quantizer 0 separates by jet flavor and leaves a fraction of codes unused, while later stages refine the kinematics.

\begin{table}[htbp!]
\centering

\caption{Background rejection ($1/\varepsilon_\mathrm{bg}$) for classifiers trained on continuous constituent features versus RVQ tokenized representations ($z_q$), for tokenizers along the bitrate ladder. 
Tokenizer configurations are labeled by their
bits per constituent, $Q\log_2 K$: 60\,bits $(K=2^{15}, Q= 4)$, 52\,bits $(2^{13}, 4)$, 40\,bits $(2^{10}, 4)$,  and 30\,bits $(2^{10}, 3)$.
The compression relative to the continuous format is shown as a multiplier.
Results are quoted at 50\% and 70\% $b$-jet signal efficiency. 
Uncertainties are estimated from bootstrapping.}

\label{tab:classification}
\small
\setlength{\tabcolsep}{6pt}
\renewcommand{\arraystretch}{1.2}

\begin{tabular}{c l r rrrr}
\toprule
 & & Continuous & \multicolumn{4}{c}{Tokenized} \\
\cmidrule(lr){3-3} \cmidrule(lr){4-7}
 & Bits/constituent & 640 & 60 & 52 & 40 & 30 \\
 & Compression      & --- & $10.7\times$ & $12.3\times$ & $16.0\times$ & $21.3\times$ \\
\midrule
\multirow{3}{*}{$\varepsilon_b = 0.50$}
 & light  & $18\,500 \pm 650$ & $15\,050 \pm 490$ & $16\,090 \pm 540$ & $11\,410 \pm 320$ & $7\,780 \pm 170$ \\
 & $c$    & $791 \pm 14$ & $659 \pm 10$ & $611 \pm 9$ & $434 \pm 6$ & $204 \pm 2$ \\
 & $\tau$ & $8\,680 \pm 800$ & $6\,360 \pm 510$ & $7\,340 \pm 630$ & $4\,480 \pm 300$ & $2\,000 \pm 90$ \\
\midrule
\multirow{3}{*}{$\varepsilon_b = 0.70$}
 & light  & $1\,370 \pm 10$ & $1\,170 \pm 10$ & $1\,100 \pm 10$ & $860 \pm 7$ & $585 \pm 4$ \\
 & $c$    & $42.0 \pm 0.2$ & $39.0 \pm 0.2$ & $38.0 \pm 0.2$ & $30.2 \pm 0.1$ & $19.1 \pm 0.1$ \\
 & $\tau$ & $451 \pm 10$ & $354 \pm 7$ & $358 \pm 7$ & $271 \pm 4$ & $143 \pm 2$ \\
\bottomrule
\end{tabular}

\end{table}
\section{Conclusions}
\label{sec:conclusions}
We treated tokenization as a rate--distortion problem and tokenized charged-particle tracks, calorimeter topoclusters, and calorimeter cells from two datasets.
The hyperparameter scans performed in this work point to four design rules for building tokenizers for HEP detector data.
First, use multiple residual quantizer stages as reconstruction improves steadily with $Q$ and a single stage tends to collapse onto a small fraction of available codes. 
Second, larger codebooks are generally better and resolution improves with $K$ across all modalities, and the cost of unused codes is small compared to the gain from finer coverage of the latent space. 
Third, the code dimension $d$ should be treated as a modality-specific tuning parameter tied to the available bit budget. 
Fourth, encode detector positions $(\eta,\phi)$ in a fixed global binning when coordinates make up a large fraction of the element’s features.
This frees the learned dimensions for modality-specific features at a cost of $\log_2{K_{\mathrm{pos}}}$ bits per coordinate, and provides a position vocabulary shared across modalities.

This work demonstrates that discrete tokens preserve much of the relevant jet-level physics at high compression.
The track and topocluster tokenizers reach sub-percent jet $p_T$ resolution at compression factors of approximately 4--11$\times$, and outperform a per-feature binning baseline in MAE at every bit rate and in jet $p_T$ resolution where the 1\% target is first met.
In COCOA, jet $p_T$ regressors on the tokens approach their continuous baselines, and a particle flow model on positions-out tokens reaches within 2\% (relative) of its continuous-input counterpart.
In JetSet, classifiers trained on the tokens retain 73--93\% of the continuous-feature background rejection at 60 bits per constituent ($11\times$ compression), with the loss growing smoothly as bits are removed.
These results show that HEP detector data can be reduced to a compact discrete vocabulary at a quantifiable cost in the observables that downstream applications depend on.
This gives emerging token-based foundation models a tokenization step with quantified trade-offs.

\section*{Acknowledgements}
M.K., J.K. and S.K. are supported by the US Department of Energy (DOE) under Grant No. DE-AC02-76SF00515, and the American Science Cloud HEP Intelligent Data Activity TREASURE project within the US DOE Genesis Mission.   
N.K. and E.G. are supported by the Minerva Grant, the Knell Family Institute for Artificial Intelligence, the  Benoziyo Center for High Energy Physics, a BSF Grant, and the Krenter-Perinot Center for High Energy Particle Physics.
L.H. is supported by BMFTR Project SciFM 05D25WO2. 

\section*{Code and data availability}
The code and workflows used to produce the results in this paper are available at \url{https://github.com/Treasure-AmSC/precision-tokenization}.
The JetSet dataset is publicly available from the CERN Open Data Portal~\cite{jetset}. The COCOA dataset is described in~\cite{COCOA}.
\bibliographystyle{elsarticle-num}
\bibliography{references}

\clearpage

\appendix
\section{Further Dataset Information}
\label{app:datasets}

\begin{table}[htbp]
\centering
\caption{Summary of the datasets and detector modalities used in this work: data provenance,
modalities, type, multiplicity, dataset size, task, and preprocessing strategy. 
JetSet includes tracks from $t\bar{t}$, while COCOA is a single-jet simulation that includes tracks, topoclusters, and calorimeter cells.
The three modalities in COCOA are extracted from the same events.
Only JetSet carries classification labels.
}
\label{tab:datasets}
\small
\setlength{\tabcolsep}{4pt}
\renewcommand{\arraystretch}{1.2}
\begin{tabularx}{\textwidth}{@{}
  >{\raggedright\arraybackslash}p{2.1cm}
  >{\raggedright\arraybackslash}p{2cm}
  l
  >{\raggedright\arraybackslash}p{1.2cm}
  >{\raggedright\arraybackslash}p{1.6cm}
  >{\raggedright\arraybackslash}X @{}}
\toprule
Dataset (gen.\ + sim.) & Modality & Element \ type & Elements\ /event & \#Jets (events) & Task / preprocessing \\
\midrule
JetSet~\cite{jetset}\newline ($t\bar{t}$, ATLAS)
  & Jet constituents (tracks) & set & $\le 40$ & $200$M ($49$M evt) & reconstruction and flavor-tagging ($\ell,c,b,\tau$); features log-standardized \\
\midrule
\multirow{3}{*}{\shortstack[l]{COCOA~\cite{COCOA}\\ (single-jet)}}
  & Tracks        & set   & $\le 15$                & \multirow{3}{*}{$90$M} & reconstruction; features log-standardized \\
  & Topoclusters & set   & $\le 50$                &                        & reconstruction; features log-standardized \\
  & Calo.\ cells  & image & $156$ patches ($64$ pixels/patch)  &                        & reconstruction; $E^{0.3}$ rescaling \\
\bottomrule
\end{tabularx}
\end{table}

\clearpage
\section{Calorimeter Cell Tokenization Procedure}
\label{app:calo}

\begin{figure}[htbp]
    \centering
    \includegraphics[width=0.99\linewidth]{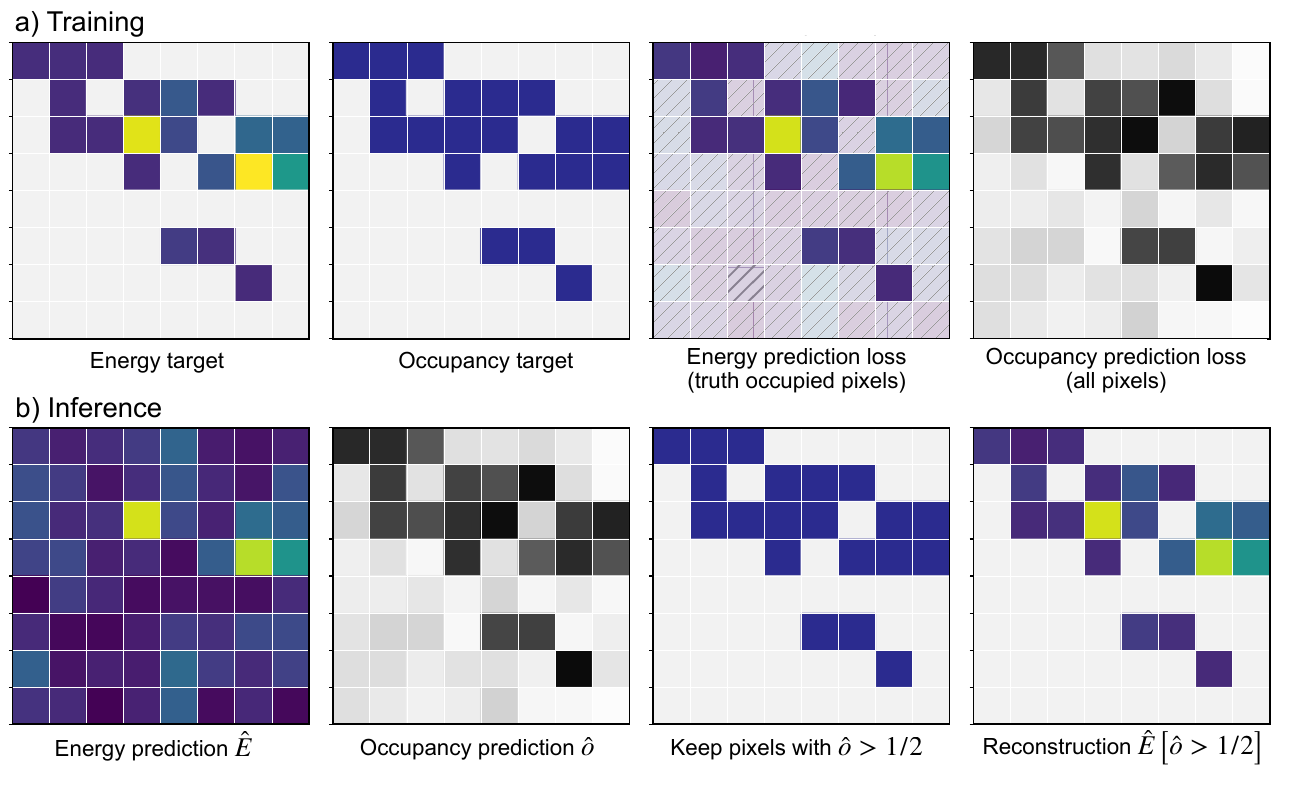}
    \caption{Calorimeter tokenization objective for a single $8\times8$ patch.
    (a) Training: the energy prediction is compared to the target only on truth-occupied pixels (hatched pixels are ignored), while the occupancy prediction $\hat{o}$ is compared to the target occupancy on all 64 pixels.
    (b) Inference: energy $\hat{E}$ and occupancy $\hat{o}$ are predicted.
    Pixels with $\hat{o}>1/2$ are kept and take their energy from $\hat{E}$.
    All others are set to zero.
}    \label{fig:placeholder}
\end{figure}

\clearpage
\section{Model and Dataset Size Studies}
\label{app:model_dataset_size}

\begin{figure}[htbp]
\centering
\includegraphics[width=0.7\textwidth]{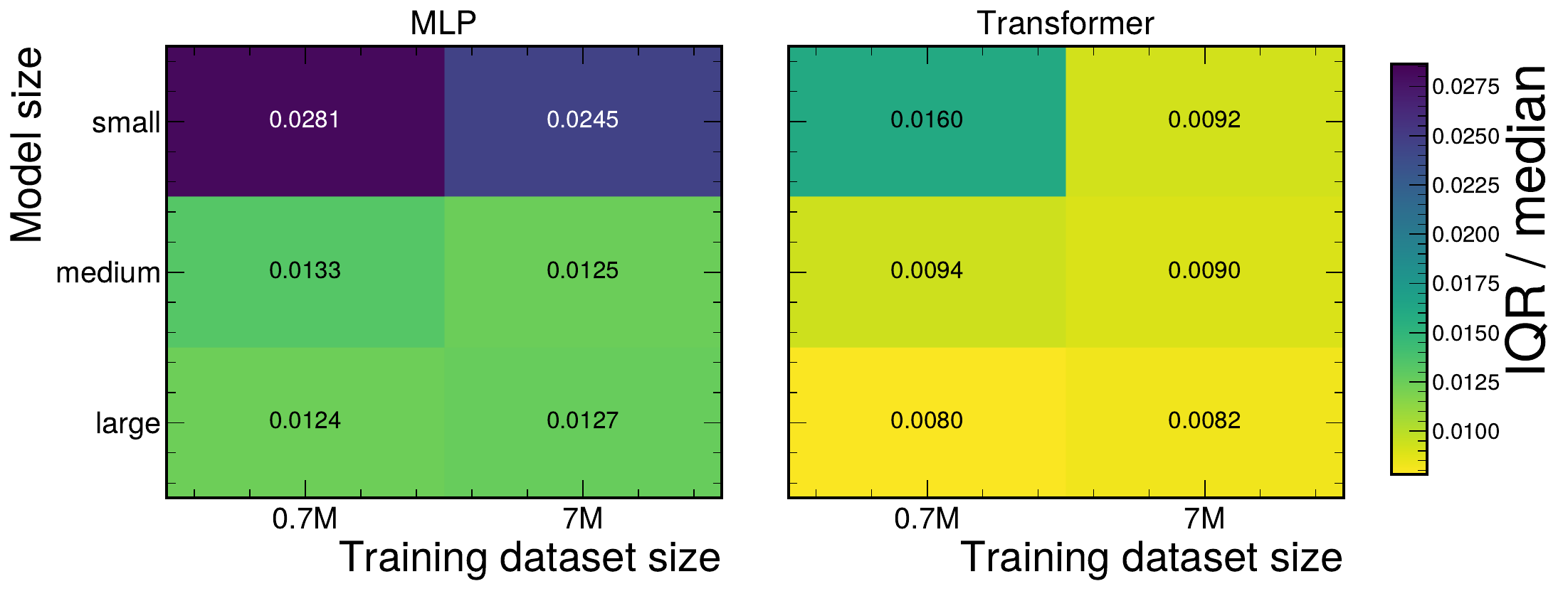}
\caption{Reconstructed jet resolution comparing no-context encoding architecture (MLP, left) against in-context encoding architecture (transformer, right).
A scan is performed over model size and training dataset size on the JetSet dataset.
The Transformer outperforms the MLP at every configuration.
}
\label{fig:mlp_vs_transformer}
\end{figure}

\begin{table}[htbp]
    \centering
    \caption{Jet resolution (IQR/median) versus training-set size, at a fixed tokenizer configuration $(K, d, Q)$ per modality.
    }
    \label{tab:datasize}
    \begin{tabular}{llrc}
    \toprule
    Modality & $(K, d, Q)$ & Training events & IQR/median [\%] \\
    \midrule
    \multirow{4}{*}{COCOA tracks}       & \multirow{4}{*}{$(4096, 4, 4)$}  & 1~M  & 0.068 \\
    &                                  & 5~M  & 0.049 \\
    &                                  & 10~M & 0.061 \\
    &                                  & 20~M & 0.049 \\
    \midrule
    \multirow{3}{*}{COCOA topoclusters} & \multirow{3}{*}{$(4096, 8, 4)$}  & 1~M  & 1.48 \\
    &                                  & 5~M  & 1.03 \\
    &                                  & 10~M & 1.01 \\
    \midrule
    \multirow{2}{*}{JetSet}             & \multirow{2}{*}{$(32768, 8, 4)$} & 0.7~M  & 0.80 \\
    &                                  & 7~M & 0.82 \\
    \midrule
    \multirow{4}{*}{COCOA cells}        & \multirow{2}{*}{$(4096, 8, 8)$}  & 1.6~M  & 3.76 \\
    &                                  & 16~M & 3.27 \\
    & \multirow{2}{*}{$(4096, 8, 16)$} & 1.6~M  & 3.62 \\
    &                                  & 16~M & 3.24 \\
    \bottomrule
    \end{tabular}
\end{table}

\clearpage  
\section{Codebook Utilization Studies}
\label{app:codebook_utilization}

\begin{figure}[htbp!]
\centering
\begin{subfigure}[b]{0.48\textwidth}
  \centering
  \includegraphics[width=\linewidth]{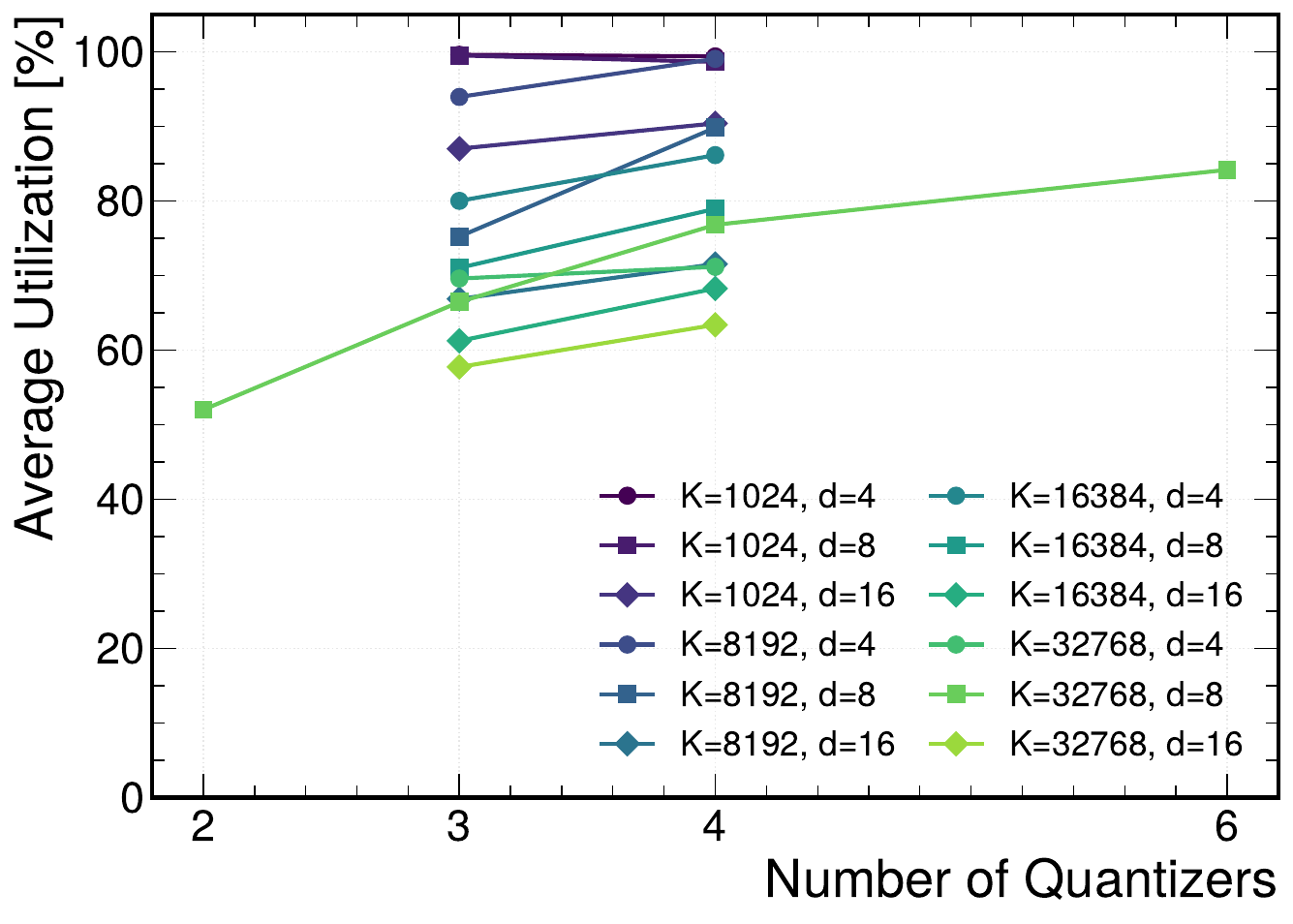}
  \caption{JetSet tracks.}
  \label{fig:jetset_utilization}
\end{subfigure}
\hfill
\begin{subfigure}[b]{0.48\textwidth}
  \centering
  \includegraphics[width=\linewidth]{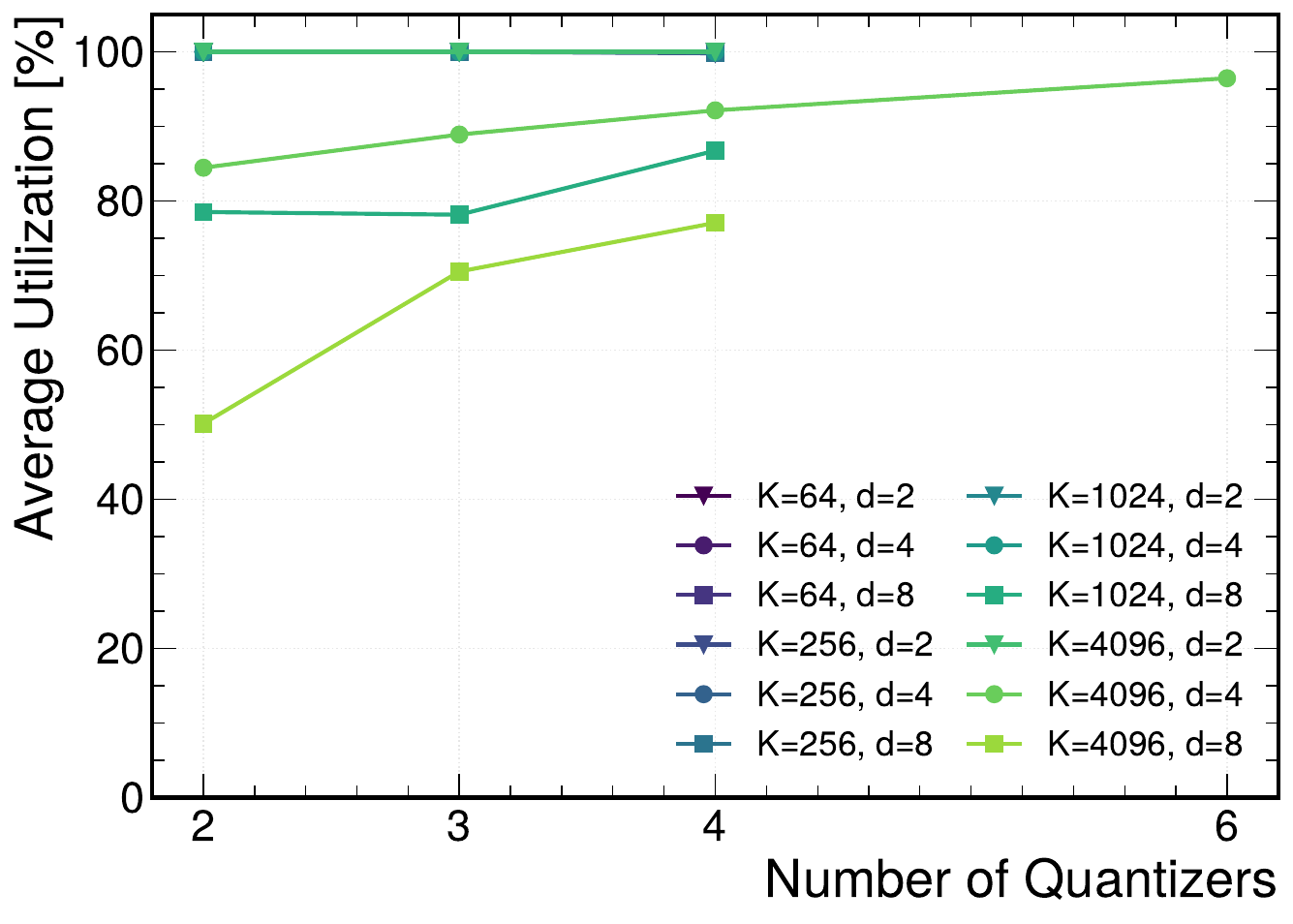}
  \caption{COCOA tracks.}
  \label{fig:cocoa_track_tokenization}
\end{subfigure}

\vspace{1em}

\begin{subfigure}[b]{0.48\textwidth}
  \centering
  \includegraphics[width=\linewidth]{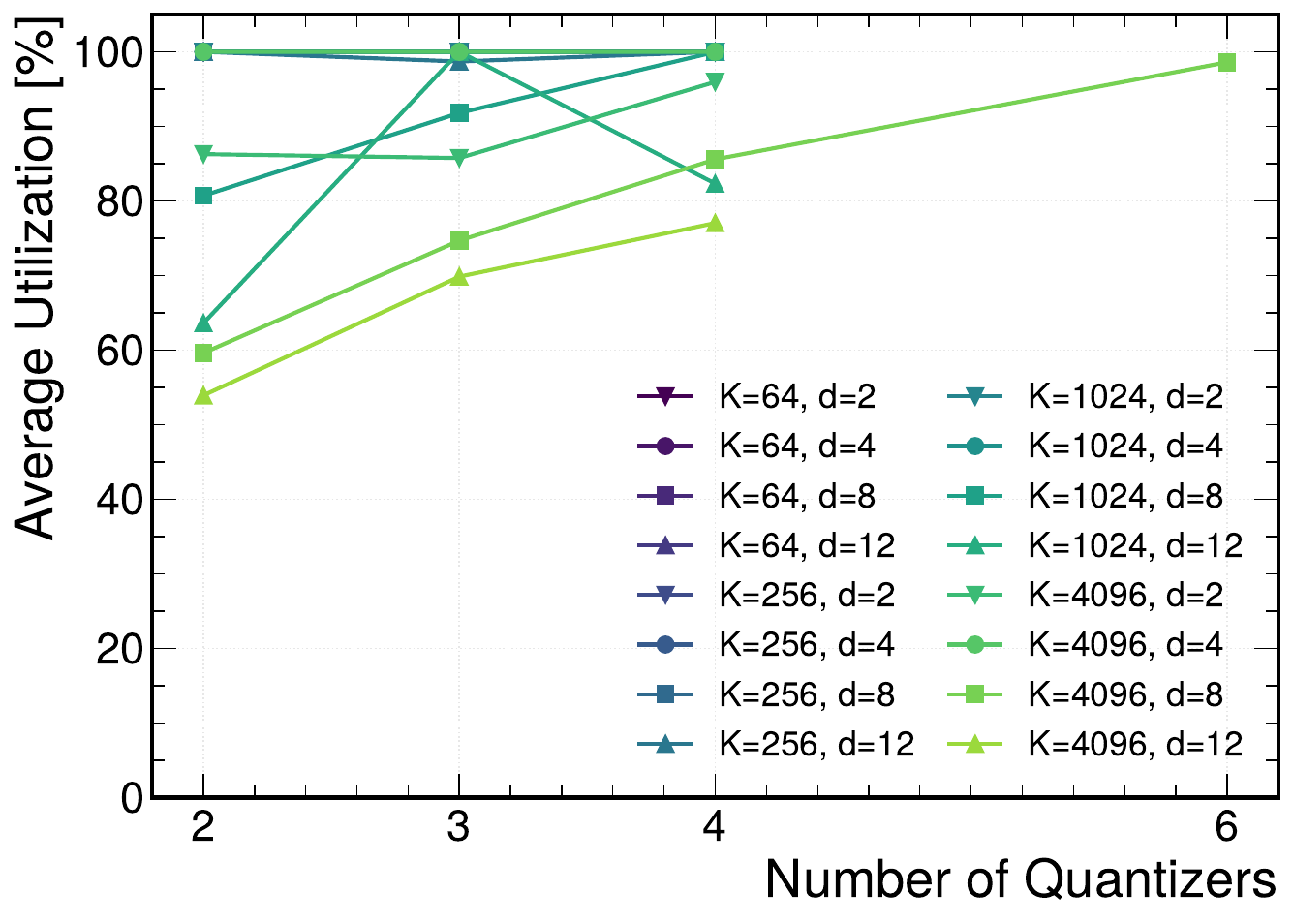}
  \caption{COCOA topoclusters.}
  \label{fig:cocoa_topo_tokenization}
\end{subfigure}
\hfill
\begin{subfigure}[b]{0.48\textwidth}
  \centering
  \includegraphics[width=\linewidth]{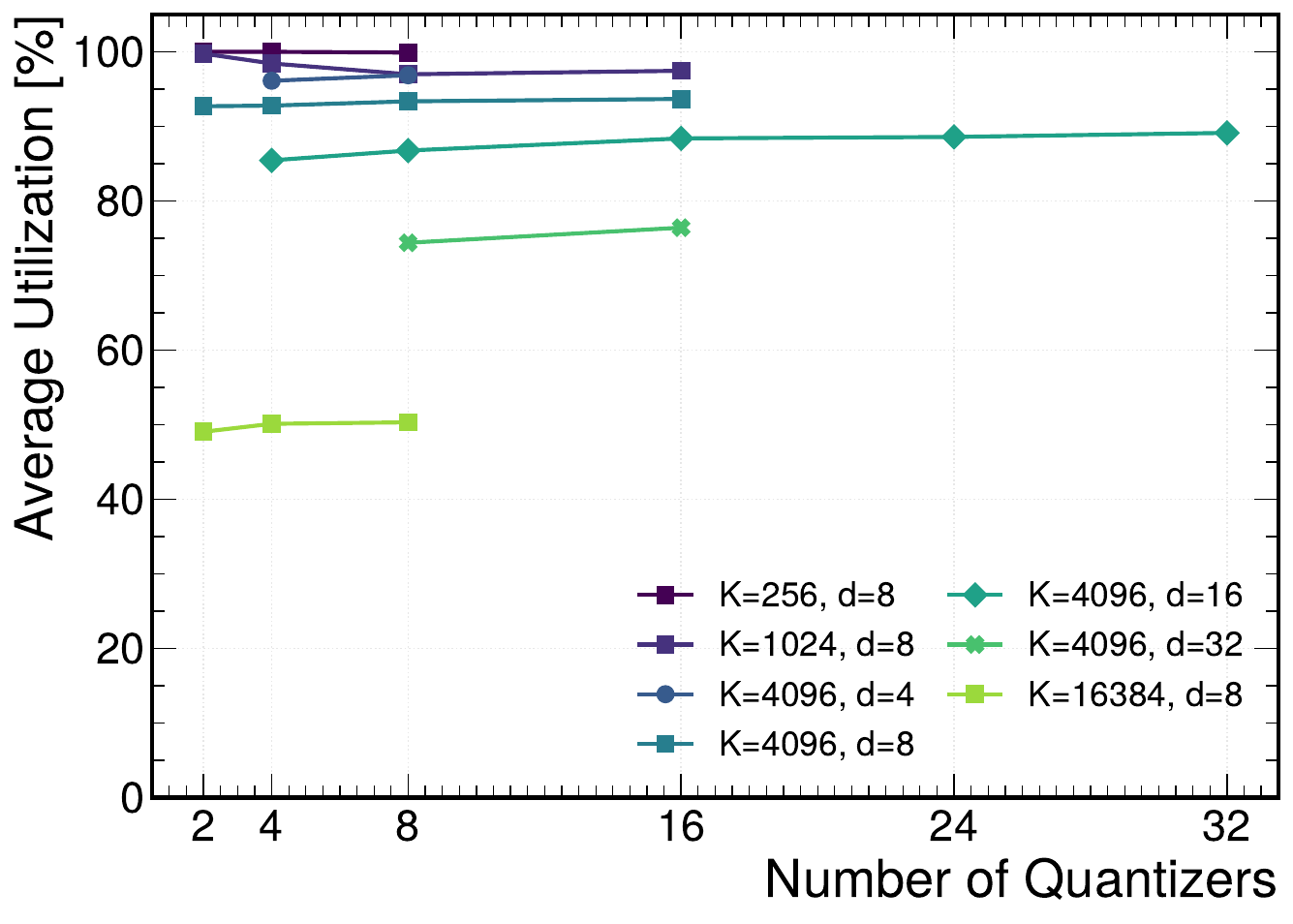}
  \caption{COCOA calorimeter cells.}
  \label{fig:cocoa_cell_tokenization}
\end{subfigure}
\caption{Average codebook utilization versus the number of quantizers $Q$, for each scanned $(K, d)$. Configurations not visible lie at 100\%.}
\label{fig:codebook_utilization_heatmaps}
\end{figure}

\clearpage
\section{Further Model Performance Studies}
\label{app:model_performance_studies}

\begin{figure}[!htbp]
    \centering
    \includegraphics[width=0.5\linewidth]{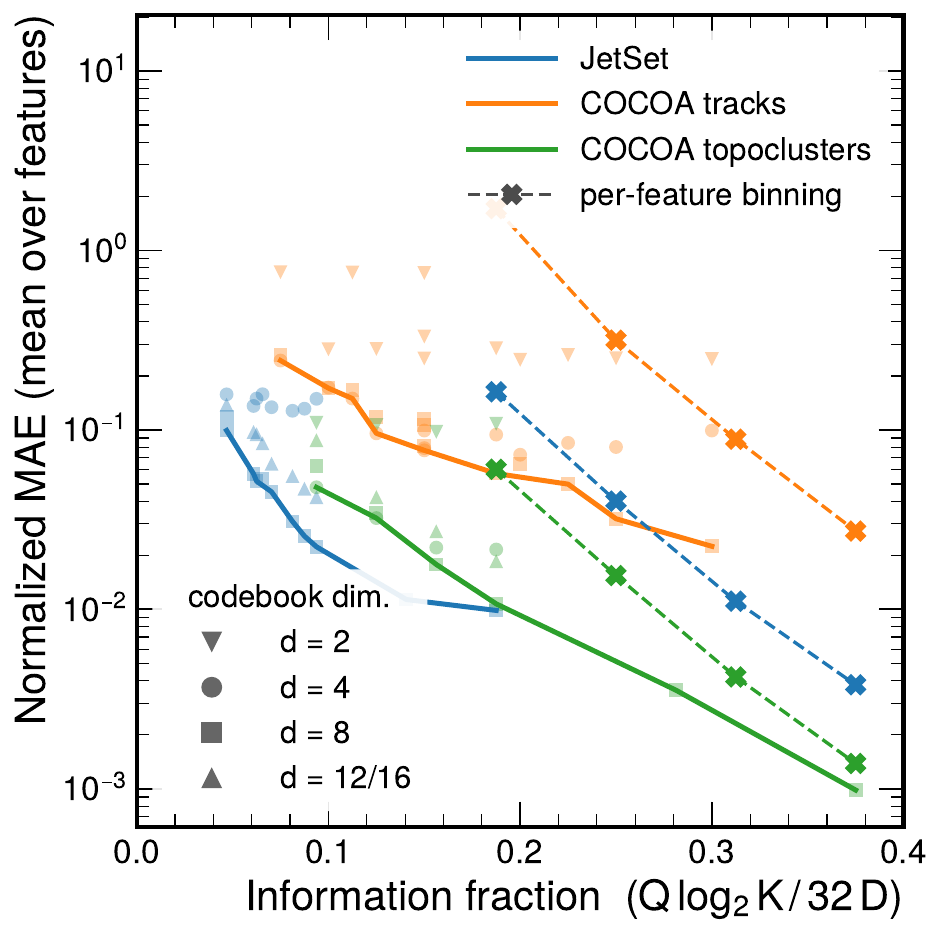}
    \caption{Pareto frontier of normalized mean absolute error (MAE)  versus information fraction $Q\log_2 K/(32D)$ for the JetSet track, COCOA track, and COCOA topocluster tokenizers.
    Marker shape denotes the codebook dimension $d$ (between 2--16).
    Faded points are individual $(K,d,Q)$ configurations, and lines trace the frontier for each tokenizer.
    Each feature's MAE is normalized by half the width of the central 68\% interval of its true distribution.
    } 
    \label{fig:pareto-optimal-mae}
    
\end{figure}

\begin{figure}[htbp]
\centering
\begin{subfigure}{\textwidth}
\includegraphics[width=\textwidth]{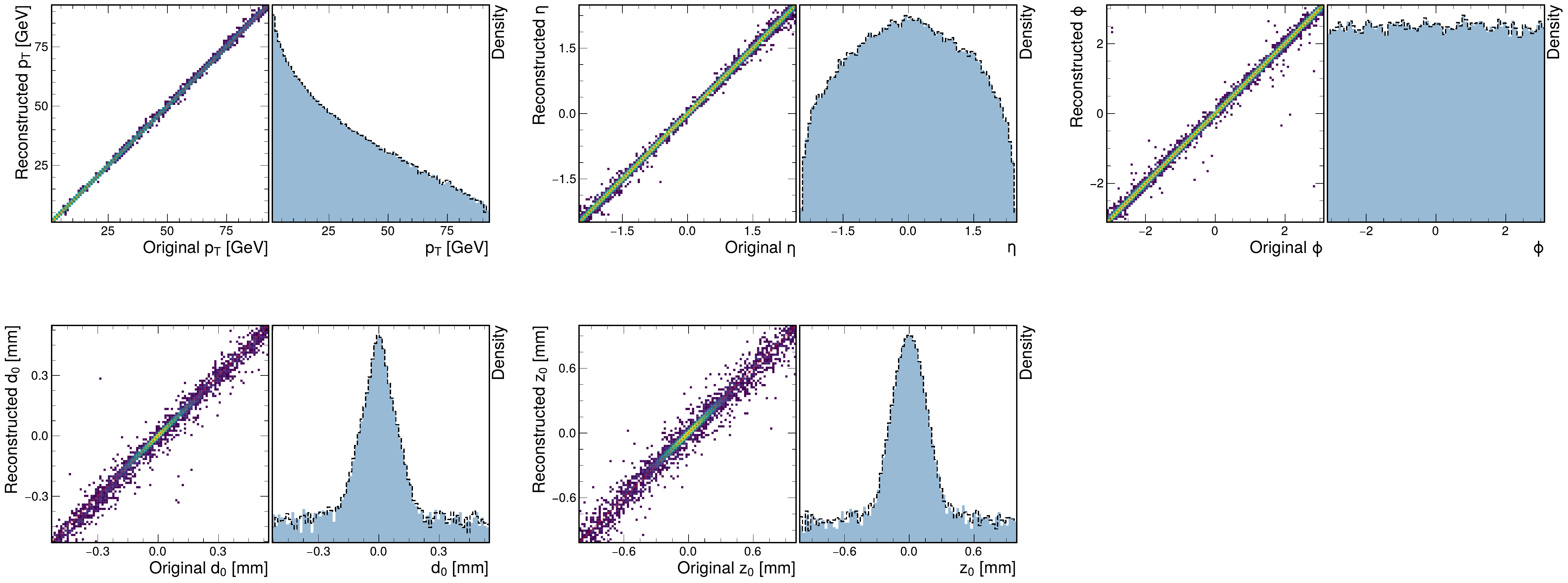}
\caption{COCOA track features for chosen RVQ with parameters $(K,d,Q)=(4096,8,3)$}
\label{fig:cocoa_track_best_withresiduals}
\end{subfigure}
\hfill
\begin{subfigure}{\textwidth}
\includegraphics[width=\textwidth]{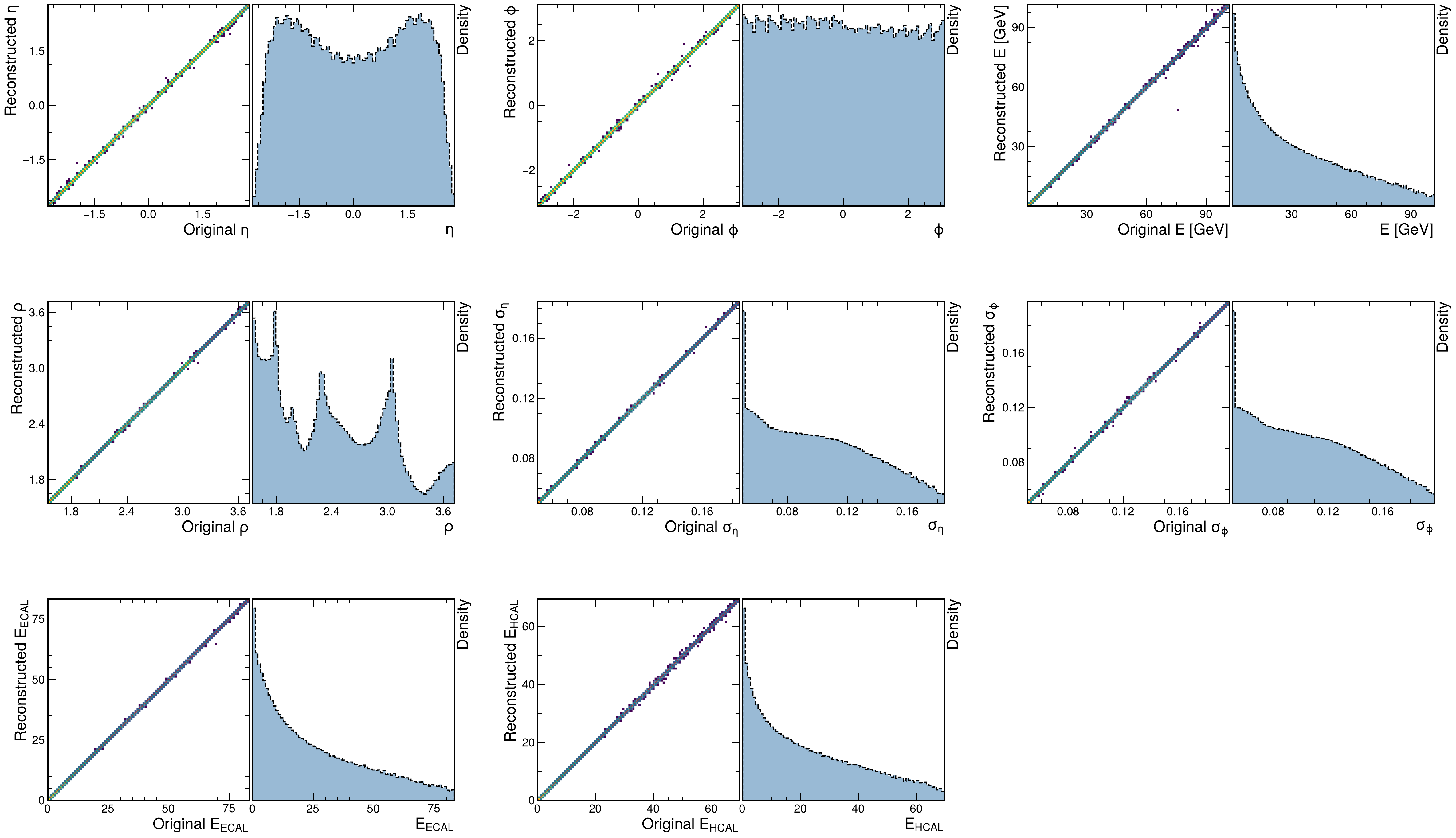}
\caption{COCOA topocluster features for chosen RVQ with parameters $(K,d,Q)=(4096,8,6)$}
\label{fig:cocoa_topos_best_withresiduals}
\end{subfigure}
\caption{Paired correlation and marginal distributions for each of the input features for the VQ-VAE for (a) the track modality and (b) the topocluster modality. 
Each pair of plots shows reconstructed-truth correlation and a marginal distribution.
In the marginal distributions, the truth distribution is shown as a dashed line, while the reconstructed is shown as a filled histogram.}
\label{fig:cocoa_per_feature}
\end{figure}

\begin{figure}[htbp]
\centering
\includegraphics[width=0.6\textwidth]{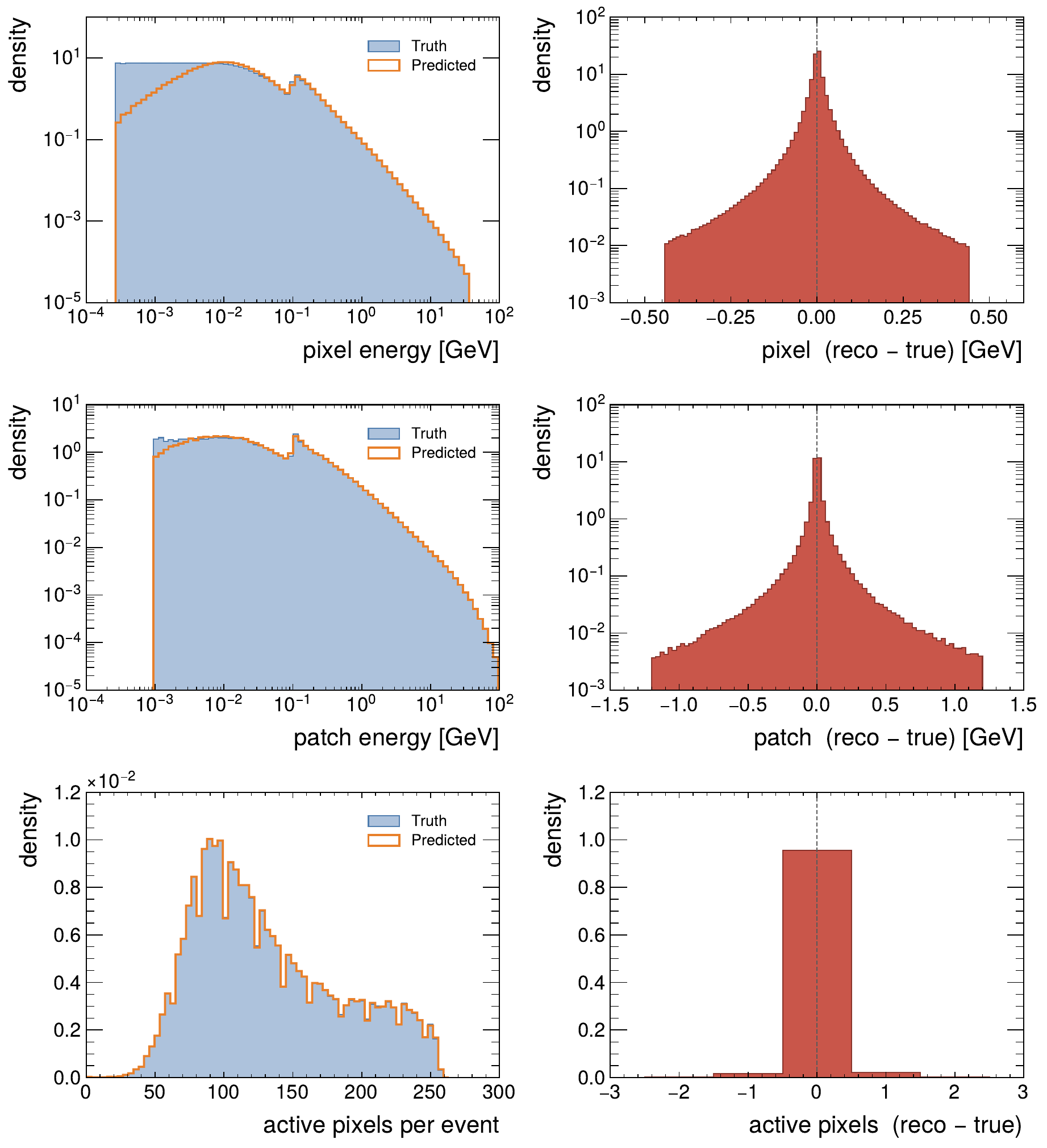}
\caption{
Energy and occupancy reconstruction for the optimal cell configuration
$(K,d,Q)=(4096,16,16)$. 
\textit{Top:} pixel energy marginals and residual $(E^\mathrm{reco}-E^\mathrm{truth})$. 
\textit{Middle:} patch energy marginals and residuals. 
\textit{Bottom:} per-event active-pixel multiplicity (mean $132.13$ truth, $132.14$ reconstructed) marginals and residuals. 
The tokenizer reproduces the patch energy spectrum and cell multiplicity at high fidelity.
}
\label{fig:cocoa_cells_diagnostics}
\end{figure}

\end{document}